\documentclass[aps,prl,superscriptaddress,twocolumn,longbibliography, floatfix]{revtex4-2}

\usepackage{graphicx}
\usepackage{amssymb}
\usepackage{amsmath}
\usepackage{hyperref}
\usepackage[utf8]{inputenc}
\usepackage{mathtools}
\usepackage[english]{babel}
\usepackage{bbm}
\hypersetup{colorlinks=true, linkcolor=blue, citecolor=blue, urlcolor=blue}
\usepackage{xcolor}
\usepackage[normalem]{ulem}
\usepackage{booktabs}
\usepackage{tabularx}

\usepackage{threeparttable}

\begin{document}

\title{Detection of HCN and diverse redox chemistry in the plume of Enceladus}

\author{Jonah S. Peter}
\email{jonahpeter@g.harvard.edu}
\affiliation{Jet Propulsion Laboratory, California Institute of Technology, Pasadena, California 91109, USA}
\affiliation{Biophysics Program, Harvard University, Boston, Massachusetts 02115, USA}
\author{Tom A. Nordheim}
\affiliation{Jet Propulsion Laboratory, California Institute of Technology, Pasadena, California 91109, USA}
\author{Kevin P. Hand}
\affiliation{Jet Propulsion Laboratory, California Institute of Technology, Pasadena, California 91109, USA}

\begin{abstract}
The \textit{Cassini} spacecraft observed that Saturn’s moon Enceladus possesses a series of jets erupting from its South Polar Terrain. Previous studies of in situ data collected by \textit{Cassini}’s Ion and Neutral Mass Spectrometer (INMS) have identified H$_2$O, CO$_2$, CH$_4$, NH$_3$, and H$_2$ within the plume of ejected material. Identification of minor species in the plume remains an ongoing challenge, owing to the large number of possible combinations that can be used to fit the INMS data. Here, we present the detection of several new compounds of strong importance to the habitability of Enceladus, including HCN, C$_2$H$_2$, C$_3$H$_6$, and C$_2$H$_6$. Our analyses of the low velocity INMS data, coupled with our detailed statistical framework, enable discrimination between previously ambiguous species in the plume by alleviating the effects of high dimensional model fitting. Together with plausible mineralogical catalysts and redox gradients derived from surface radiolysis, these compounds could potentially support extant microbial communities or drive complex organic synthesis leading to the origin of life.
\end{abstract}

\maketitle

Shortly after its arrival in the Saturn system, the \textit{Cassini} spacecraft detected intense plume activity at the mid-sized moon, Enceladus~\cite{dougherty_identification_2006, porco_cassini_2006}. \textit{Cassini} in situ and remote sensing observations have confirmed that the plume consists primarily of H$_2$O gas~\cite{hansen_water_2008, waite_cassini_2017, waite_jr_liquid_2009, waite_cassini_2006} as well as H$_2$O-ice grains that feed Saturn’s E-ring~\cite{hillier_composition_2007, postberg_salt-water_2011, postberg_sodium_2009, postberg_e-ring_2008}. CO$_2$ has also been detected in both the gaseous phase within the plume itself~\cite{waite_cassini_2017, waite_jr_liquid_2009, waite_cassini_2006}  and as a condensate in plume deposits on Enceladus’ surface~\cite{combe_nature_2019}. In situ measurements of the plume’s neutral gas component made by \textit{Cassini}’s Ion and Neutral Mass Spectrometer (INMS) further indicate the presence of CH$_4$, NH$_3$, and H$_2$ within Enceladus’ subsurface ocean~\cite{waite_cassini_2017}. Although early publications identified several additional species within the plume~\cite{waite_jr_liquid_2009}, more recent work suggests that many of these compounds resulted from incidental high velocity impact fragmentation of larger molecules within the instrument antechamber~\cite{postberg_macromolecular_2018, waite_cassini_2017}. Analyses of plume material sampled during the lower velocity flybys of Enceladus (for which this fragmentation was less significant) do imply the existence of additional plume species, however no study to date has been able to verify the identity of any other intrinsic compounds.

\begin{figure*}[t]
\centering 
\renewcommand\figurename{Fig.}
\includegraphics[width=\textwidth]{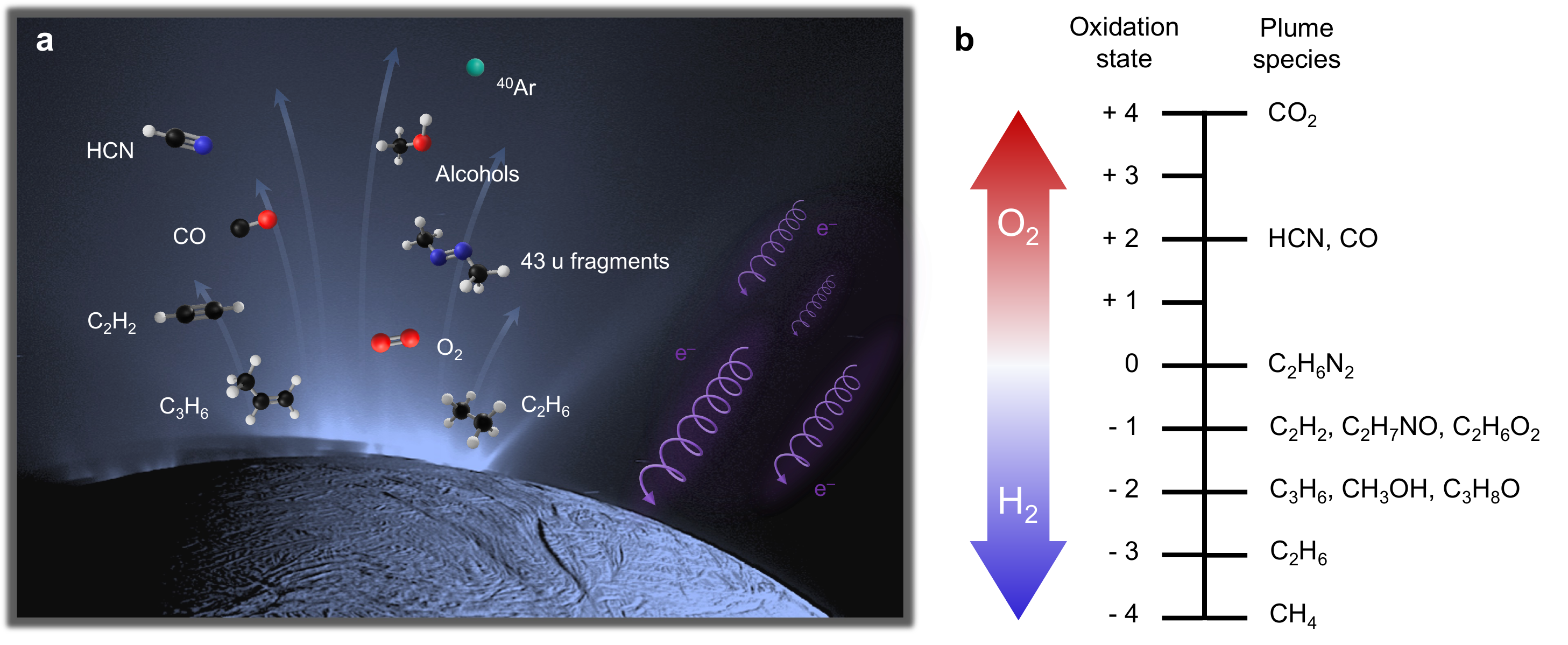}
\caption{\label{fig:overview}New compounds identified in the Enceladus plume indicate a potentially habitable environment. (a) Jets emanating from ice fissures in Enceladus' South Polar Terrain feed a plume of ejected material containing organic molecules with varying oxidation states. Electron bombardment of the surface might help facilitate the production of oxidants and prebiotic feedstock molecules observed in the plume. These compounds could potentially support biologically-mediated redox metabolisms or polymerize to form nucleic and amino acid precursors leading to the origin of life. (b) The average oxidation state of carbon for organic compounds confirmed or suspected in the plume. Plume-derived H$_2$ and O$_2$ could act as strong reducing and oxidizing agents, respectively, and may be responsible for the diverse redox chemistry seen at Enceladus.}
\end{figure*}

Difficulty in resolving minor plume constituents stems from the large number of plausible compounds relative to the low mass resolution of INMS. When training statistical models in this high dimensional regime, simple regression techniques tend to form overly complex models that produce specious results~\cite{james_introduction_2013, chandrasekaran_quantization_1974, hughes_mean_1968, trunk_problem_1979}. Models of INMS spectra suffer from an additional complexity in that the signals produced by individual molecules are not necessarily linearly independent. As such, there may be multiple different combinations of species that appear to fit the data equally well. The resulting large correlations between model components generally reduce model performance and can mask the importance of any particular component by limiting statistical power~\cite{guyon_introduction_2003, james_introduction_2013}. Studies that leave these issues unaddressed are likely to encounter model ambiguities that preclude reliable statistical inference about the plume’s composition. 

In this work, we seek to resolve the apparent compositional ambiguity of Enceladus’ plume. By characterizing the information content of the average low velocity spectrum obtained during the E14, E17, and E18 flybys, we determine constraints on the number of species that can reliably be extracted from the INMS dataset under spacecraft conditions that were least modified by fragmentation. We then use relative entropy minimization to assess the likelihood of tens of billions of potential models and show that multi-model inference allows for the identification of several new compounds not previously confirmed at Enceladus. Our results indicate the presence of a rich, chemically diverse environment that could support complex organic synthesis and possibly even the origin of life (Fig.~\ref{fig:overview}).

\section{Constraints on the complexity of plume models}

Deconvolving the overlapping signals of each species in the plume requires comparing features in the INMS flight data to a library of known mass spectra. Consequently, species in the plume cannot be identified unless explicitly included within the models used for comparison. Previous studies have constructed model INMS spectra using a variety of methods, including singular value decomposition~\cite{cui_analysis_2009} and the application of custom-defined fit statistics~\cite{magee_inms-derived_2009, waite_ion_2005, waite_process_2007, waite_chemical_2018} (see Supplementary Information). In all cases, these fitting procedures seek to minimize the training error associated with the residual counts between the INMS spectrum and the reconstructed model fit. Crucially, however, the training error is not an appropriate metric for evaluating model performance. In fact, it is a well-known concept in statistical modeling that the training error will continue to decrease with the inclusion of additional model components, regardless of whether those components are genuinely related to the observed data. Instead, it is necessary to approximate how each model would perform on a set of independent observations. This is the preferred method of model validation when additional data sets are unavailable for explicit model testing~\cite{james_introduction_2013}. Analysis techniques that utilize only the training error risk developing overly complex models and claiming false species detections.

\begin{figure*}[t]
\centering 
\renewcommand\figurename{Fig.}
\includegraphics[width=\textwidth]{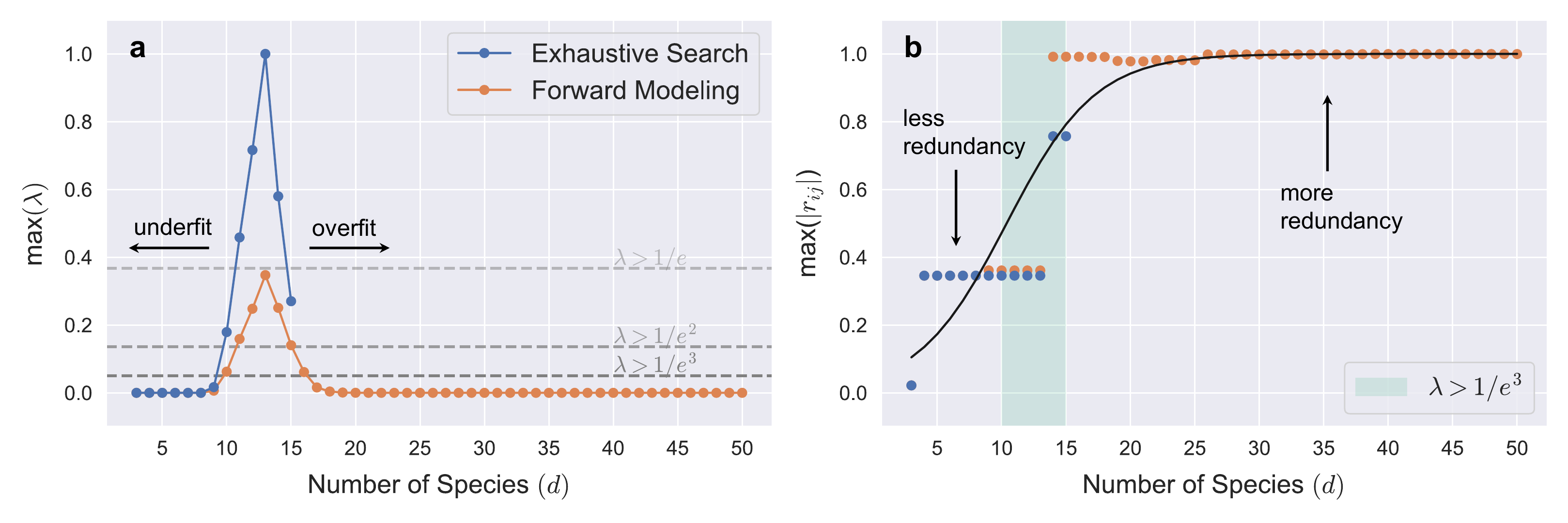}
\caption{\label{fig:complex}Model performance as a function of model complexity. (a) Maximum relative likelihoods across all models with $d$ species. Blue circles indicate best-fit (highest $\lambda$) models identified via an exhaustive search for $d<16$. Orange circles represent models constructed using a benchmark forward modeling procedure extending to $d=50$ (see Methods). There is good agreement between the trends observed with forward modeling and the exhaustive search. Models with $\lambda > 1/e^3$, $1/e^2$, and $1/e$ (dashed lines) exhibit increasingly strong predictive power. All models with $d<10$ species underfit the data, while those with $d>15$ overfit. (b) The maximum magnitude of pairwise correlations $r_{ij} = \mathrm{cov}(\beta_i \beta_j)/\sqrt{\mathrm{var}(\beta_i)\mathrm{var}(\beta_j)}$ between regression coefficients in the best-fitting models at each value of $d$. Blue and orange circles represent the same models as in panel (a). Correlations loosely follow a logistic curve demonstrating model redundancy for $d > 15$. The inflection point occurs within the optimal performance interval $d \in [10,15]$ characterized by $\lambda > 1/e^3$ (green shaded area).}
\end{figure*}

Here, we evaluate model performance using the small sample bias-corrected Akaike Information Criterion (AICc)~\cite{akaike_information_1973, hurvich_regression_1989, sakamoto_akaike_1986, sugiura_further_1978, burnham_model_2004} which estimates the relative entropy between a given model and the unknown, true distribution that produced the observed data. We construct each candidate model as a linear combination of end-member spectra representing the different cracking patterns of individual molecules. A model $M$ is given by
\begin{equation}\label{eq1}
    M: \quad \hat{y}_i = \sum^d_{k=1} \beta_k x_{k,i},
\end{equation}
where $\hat{y}_i$ is the total modeled counts in mass channel $i$, $x_{k,i}$ is the cracking pattern of species $k$ at mass channel $i$, $\beta_k \geq 0$ is the regression coefficient for species $k$, and $d$ is the total number of species in the model. The observed INMS counts at each mass channel, $y_i$, are treated as independent data points with unequal Gaussian uncertainties, $\sigma_i$ (see Methods). The AICc for each model is then given by
\begin{equation}\label{eq2}
    \text{AICc} = 2d - 2 \ln[\mathcal{L}(M_0 \vert y)] + \frac{2d(d+1)}{n-d-1},
\end{equation}
where $n$ is the total number of mass channels and $\mathcal{L}(M \vert y)$ is the model's likelihood function,
\begin{equation}\label{eq3}
    \mathcal{L}(M \vert y) = \prod^n_{j=1} \left(\frac{1}{\sigma_j \sqrt{2\pi}}\right) \exp \left\{- \frac{1}{2} \sum^n_{i=1}\left(\frac{y_i - \hat{y}_i}{\sigma_i}\right)^2\right\},
\end{equation}
evaluated at the maximum likelihood estimate, $M_0$, obtained by optimizing the set of $\{\beta_k\}$. The last term in Eq.~\eqref{eq2} is a correction factor that penalizes overly complex models when the sample size is small ($n/d \lesssim 40$)~\cite{burnham_model_2004}. The model with the minimium AICc (AICc$_\text{min}$) asymptotically approximates the model with the lowest Kullback-Leibler information loss relative to the observed data~\cite{akaike_likelihood_1978, kullback_information_1951, burnham_model_2004}. Whereas standard maximum likelihood estimation seeks to minimize the training error quantified by the variance-weighted sum of squared residuals, $\sum^n_{i=1}((y_i - \hat{y}_i)/\sigma_i)^2$, minimization of the AICc accounts for the bias introduced by estimating the regression coefficients via the training data. The relative likelihood, or evidence ratio, of each model follows as~\cite{johnson_model_2004}
\begin{equation}
    \lambda = \exp \left\{-(\text{AICc} - \text{AICc}_{\text{min}})/2 \right\}
\end{equation}
and can be used to compare models of differing complexity. Generally, models with $\lambda < 1/e^3 \approx 0.05$ exhibit little to no predictive power while those above this threshold form a $\sim$95--99\% model confidence interval suitable for multi-model inference~\cite{burnham_model_2004, richards_dealing_2007, richards_model_2011, richards_testing_2005}.  

To capture the full range of possible plume constituents, we composed a large spectral library containing the most recently published list of plausible INMS-detectable plume species~\cite{postberg_plume_2018} as well as several additional compounds found in organic synthesis and laboratory experiments simulating icy satellites (Extended Data Table~\ref{tab:library}). Using this library, we performed an exhaustive search up to $d=15$ of tens of billions of potential models for the plume’s composition. Fig.~\ref{fig:complex}a demonstrates that optimal model performance is achieved with only 10--15 species. The maximum relative likelihood across all models peaks sharply at $d=13$ and drops precipitously for more complex models. A benchmark forward modeling procedure extending to $d=50$ confirms this trend for large $d$ (see Methods). Models with greater than 15 species drastically overfit the INMS data and incorrectly extract false signal from statistical noise. Large correlations $r_{ij} = \mathrm{cov}(\beta_i \beta_j)/\sqrt{\mathrm{var}(\beta_i)\mathrm{var}(\beta_j)}$ between regression coefficients in these models indicate the erroneous inclusion of redundant species with similar mass spectra (Fig.~\ref{fig:complex}b). This leads to overfitting and poor model performance, despite a monotonic decrease in the training error with increasing complexity. By contrast, significantly simpler models exhibit low correlations between model parameters, indicating the presence of additional features within the INMS spectrum that have not yet been fit by a candidate species (i.e., underfitting). Optimal performance occurs near the inflection point after all major data features have been explained but before redundant species are incorporated. 

\section{New species detected in the plume}

\begin{table*}[t]
\centering
\begin{threeparttable}
\caption{\label{tab:mixing}Volume mixing ratios for the Enceladus plume. Probabilities and mixing ratios are shown for all compounds with probability $>0.1$. Evidence for the remaining library species is poor (Extended Data Table~\ref{tab:full_data}). Values are calculated based on the exhaustive set of 26332 high likelihood ($\lambda > 1/e^3$) models using a multi-model averaging procedure that incorporates uncertainties due to model ambiguities (see Methods). Mixing ratios are given as mean $\pm$ SE and are scaled to incorporate the 0.9\% H$_2$ number abundance reported in ref.~\cite{waite_cassini_2017}. Upper limits (mean + 3 SE) are reported for species with mean values below their corresponding SE. Species listed under ``Confirmed" are detected with $> 2$ SE precision and represent the most conservative model of the plume. Species listed under ``Strong" are very likely detected but with mixing ratios that are less certain. Species listed in brackets are included in the alcohol mixing ratio. Species listed in parentheses are included in the 43 u fragment mixing ratio.}
\begin{tabularx}{\textwidth}{lllll}
\toprule
Evidence & Species & Probability\tnote{e}  & Mixing Ratio (\%)\tnote{e} & Previous Limit (\%)\tnote{f} \\
\midrule
Confirmed & H$_2$O                  & 1    & $95.9 \pm 0.3$       & 96--99  \\
          & CO$_2$                  & 1    & $0.45 \pm 0.04$      & 0.3--0.8  \\
          & CO\tnote{a}             & 1    & $0.72 \pm 0.07$      & 0.01--0.2, $< 0.05$\tnote{g}\tnote{h} \\
          & CH$_4$                  & 1    & $0.11 \pm 0.02$      & 0.1--0.3  \\
          & NH$_3$                  & 1    & $1.8 \pm 0.1$        & 0.4--1.3 \\
          & HCN                     & 1    & $0.11 \pm 0.02$      & 0.01--0.2 \\
          & C$_2$H$_2$              & 0.92 & $0.023 \pm 0.005$    & 0.01--0.2 \\
          & C$_3$H$_6$              & 0.87 & $0.004 \pm 0.002$  & $< 0.01$\\
          & H$_2$\tnote{b}          & $-$  & $-$                  & 0.4--1.4, 0.9\tnote{g}\\
Strong    & C$_2$H$_6$              & 0.70 & $0.013 \pm 0.010$    & 0.01--0.2\\
          & Alcohols\tnote{c}       & 0.82 & $< 0.02$             & $-$ \\
          & [CH$_3$OH]              & 0.44 & $< 0.005$            & $< 0.01$ \\
          & [C$_2$H$_7$NO]          & 0.23 & $< 0.002$            & $< 0.01$ \\
          & [C$_2$H$_6$O$_2$]       & 0.20 & $< 0.01$             & $< 0.01$ \\
          & [C$_3$H$_8$O]           & 0.11 & $< 0.005$            & $< 0.01$ \\
          & 43 u fragments\tnote{d} & 0.72 & $< 0.01$             & $-$ \\
          & (C$_2$H$_6$N$_2$)       & 0.24 & $< 0.002$            & $< 0.01$ \\
          & (C$_3$H$_6$O)           & 0.17 & $< 0.001$            & $< 0.01$ \\
          & O$_2$                   & 0.64 & $< 0.008$            & $< 0.01$, $< 0.004$\tnote{g}\tnote{h} \\
          & $^{40}$Ar               & 0.58 & $< 0.004$            & $< 0.01$ \\
Moderate  & H$_2$S                  & 0.37 & $< 0.003$            & $< 0.01$ \\
          & PH$_3$                  & 0.24 & $< 0.002$            & $< 0.01$ \\
          & C$_3$H$_8$              & 0.23 & $< 0.005$            & $< 0.01$ \\
          & C$_3$H$_5$Cl            & 0.23 & $< 0.001$            & $< 0.01$ \\
Poor      & CH$_3$CN                & 0.16 & $< 0.003$            & $< 0.01$ \\
          & C$_3$H$_4$              & 0.16 & $< 0.002$            & $< 0.01$ \\
          & CH$_5$N                 & 0.14 & $< 0.003$            & $< 0.01$ \\
\bottomrule
\end{tabularx}
\begin{tablenotes}
\item[a]{The vast majority of CO is likely produced through incidental impact fragmentation.}
\item[b]{Determination of the H$_2$ mixing ratio requires analysis of additional INMS data not examined here (see Methods).}
\item[c]{C$_4$H$_{10}$O might also contribute to this mixing ratio (see Supplementary Information).} 
\item[d]{Other mass 43 fragments produced by C$_4$H$_{10}$, C$_4$H$_6$O$_2$, C$_4$H$_9$N, C$_5$H$_9$N, C$_8$H$_{18}$, or C$_5$H$_{12}$ might also contribute to this mixing ratio.}
\item[e]{Denotes values calculated in this work.}
\item[f]{Denotes values presented in ref.~\cite{postberg_plume_2018} unless otherwise specified.}
\item[g]{Denotes values presented in ref.~\cite{waite_cassini_2017}.}
\item[h]{Denotes number density relative to H$_2$O.}
\end{tablenotes}
\end{threeparttable}
\end{table*}

The most recently published list of neutral gas species confirmed in the plume consists only of H$_2$O, CO$_2$, CH$_4$, NH$_3$, and H$_2$~\cite{postberg_plume_2018, waite_cassini_2017}. According to that work, no other compounds could be definitively detected by INMS during the low velocity flybys due to model ambiguities at low mixing ratios. Here, we account for those ambiguities via a multi-model averaging procedure that treats the minimum Kullback-Leibler divergence as a statistical random variable and weights each model by its probability of minimizing this information loss (see Methods). Whereas previous studies have relied on ad-hoc assessments of model ambiguity, the procedure here is rooted in the fundamental concepts of information theory and explicitly incorporates uncertainties resulting from the modeling process into the standard error (SE) of each mixing ratio. 

\begin{figure}[t]
\centering 
\renewcommand\figurename{Fig.}
\includegraphics[width=\columnwidth]{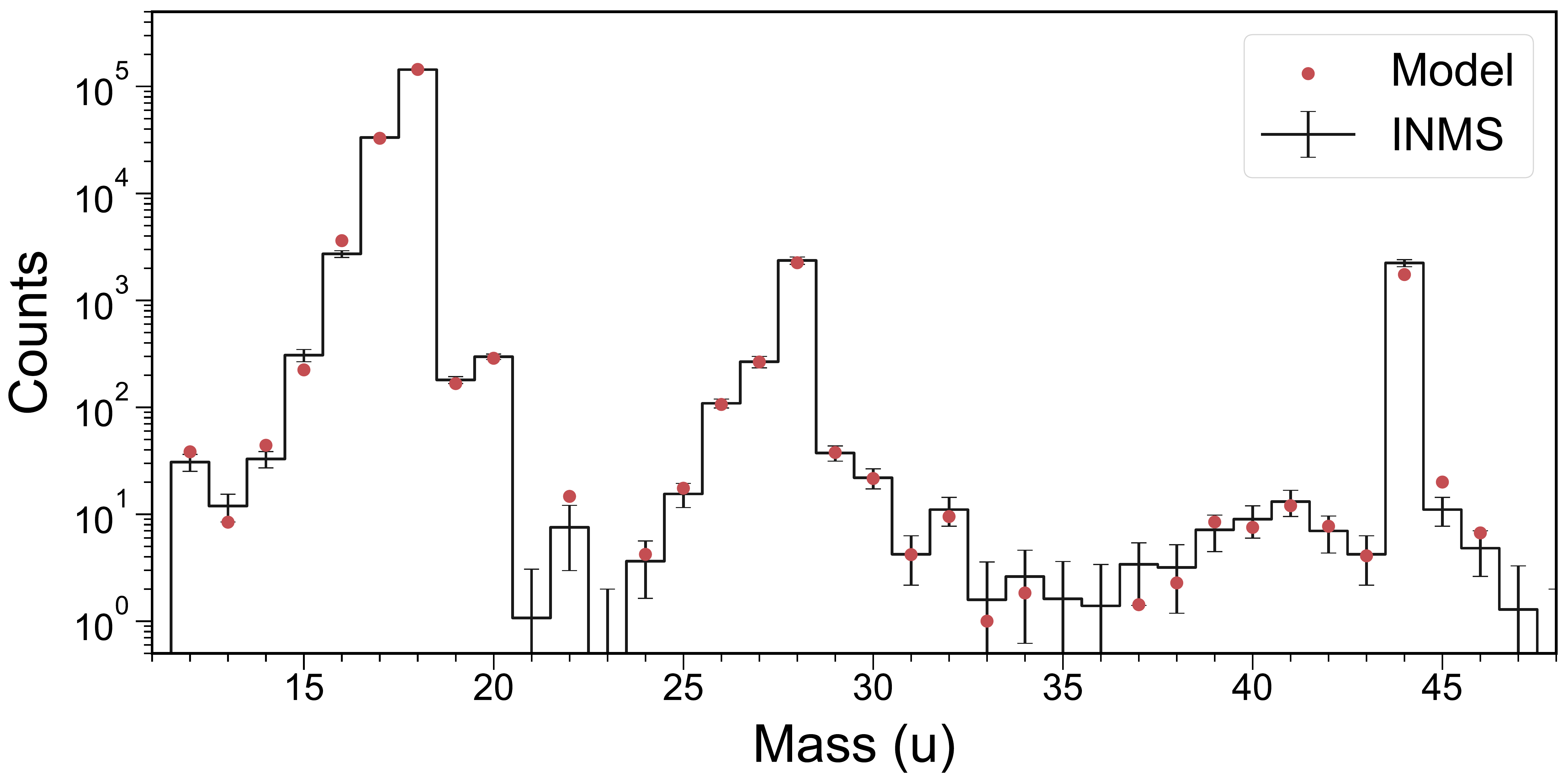}
\caption{\label{fig:spectrum}Average low velocity INMS spectrum and reconstructed model fit. The black silhouette shows the 12-47 u range of the average INMS spectrum obtained during the E14, E17, and E18 flybys. Data was adapted from ref.~\cite{postberg_macromolecular_2018}. Counts at all other mass channels are at or below the minimum count uncertainty. Error bars show 1$\sigma$ Gaussian uncertainty in the observed counts. Red circles indicate the model fit based on the mixing ratios in Table~\ref{tab:mixing}.}
\end{figure}

In addition to H$_2$O, CO$_2$, CH$_4$, and NH$_3$, we find significant evidence—beyond 2 SE precision—for HCN, C$_2$H$_2$, C$_3$H$_6$, and CO in the plume (Table~\ref{tab:mixing}). Our results are agnostic to the presence of H$_2$ which requires analysis of additional INMS data not examined here (see Methods). We also demonstrate strong evidence for C$_2$H$_6$ at 1 SE precision, as well as the very probable detection of an alcohol (most likely CH$_3$OH), molecular O$_2$, $^{40}$Ar, and an unidentified 43 u fragment (possibly from C$_2$H$_6$N$_2$ or C$_3$H$_6$O). Aside from CO, which is known to be produced as a fragmentation product within the INMS antechamber~\cite{waite_jr_liquid_2009, waite_cassini_2017, postberg_macromolecular_2018, jaramillo-botero_hypervelocity_2012}, these species are most likely intrinsic to the plume itself. In general, our mixing ratios are in excellent agreement with the most recently published limits~\cite{postberg_plume_2018, waite_cassini_2017}. The reconstructed model fit is plotted alongside the INMS spectrum in Fig.~\ref{fig:spectrum}.

Interestingly, the NH$_3$ abundance derived here is somewhat higher than that reported by Waite et al.~\cite{waite_cassini_2017} and leaves a large negative residual at 16 u (Extended Data Fig.~\ref{fig:full_spectrum}). This larger mixing ratio stems from the residual signal at masses 14--16 left by fitting H$_2$O, CO$_2$, and CH$_4$ (Fig.~\ref{fig:compare}a). However, we note that masses 17 and 18 are predominantly dictated by H$_2$O, which is used to standardize the count rates across the E14, E17, and E18 flybys (see Methods). As such, the small count uncertainties at these mass channels are likely an underestimate of the true signal variability at Enceladus and may artificially enhance the NH$_3$ mixing ratio (see Supplementary Information). Chemisorption of NH$_3$ molecules collected during previous flybys onto the INMS instrument antechamber could also contribute to elevated count rates at these mass channels~\cite{jaramillo-botero_hypervelocity_2012}. Nevertheless, some amount of NH$_3$ is likely required to explain the presence of nitrogen-bearing ions observed in Saturn’s inner magnetosphere, consistent with a source from Enceladus~\cite{smith_enceladus_2008}.

\begin{figure*}[htbp]
\centering 
\renewcommand\figurename{Extended Data Fig.}
\includegraphics[width=0.7\textwidth]{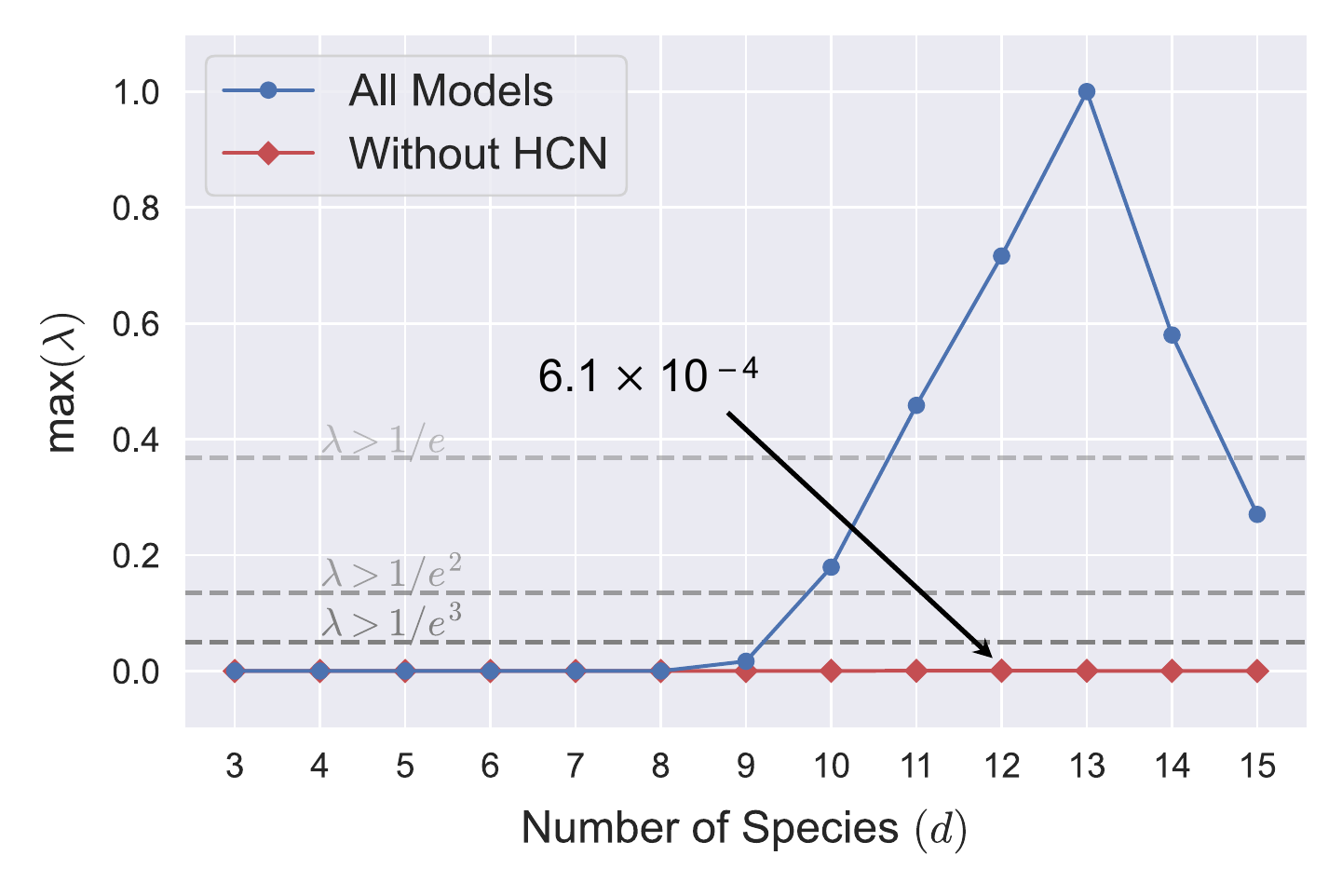}
\caption{\label{fig:complexity_HCN}Comparison of model performance without HCN. Blue circles show the maximum relative likelihoods across all models for each value of $d$ species (as in Fig.~\ref{fig:complex}a). Red diamonds indicate the highest likelihood models without HCN. All models without HCN exhibit exceptionally poor performance, with a maximum relative likelihood peaking at $\lambda = 6.1 \times 10^{-4}$ for $d = 12$.}
\end{figure*}

A major finding of this work is that nitrogen is also definitively present at Enceladus in the form of HCN. Previous studies have been unable to resolve the HCN abundance due to confounding signals from fragmentation products at mass 28. In their analysis of the E2 flyby, Waite et al.~\cite{waite_cassini_2006} quote an upper limit of $< 0.5$\%, whereas upper limits of $< 0.58$\% and $< 0.74$\% are given by Waite et al.~\cite{waite_jr_liquid_2009}, based on the high velocity E3 and E5 flybys. For the faster flybys in particular, the authors note that the elevated count rate at 28 u introduces a model ambiguity at masses 27 and 28 (N$_2$+HCN versus C$_2$H$_4$) that precludes the identification of HCN. By contrast, our procedure allows for the unambiguous identification of CO as the major 28 u fragment (CO is present in all high likelihood models---see Extended Data Table~\ref{tab:full_data}). This conclusion is strongly supported by the fragmentation signatures observed by INMS at higher velocities, and by INMS ice grain spectra~\cite{postberg_macromolecular_2018}. Properly accounting for the CO signal (and the minor contribution from C$_2$H$_6$) at mass 28 eliminates alternative model fits and reveals that HCN is required at 27 u (see Fig.~\ref{fig:compare}b and Extended Data Fig.~\ref{fig:complexity_HCN}). The mixing ratio reported here ($0.11 \pm 0.02\%$) is consistent with that observed in cometary comae~\cite{bockelee-morvan_comets_2004, mumma_chemical_2011, newburn_comets_1991}.

We also report the first strong evidence for native alcohols and additional evidence for O$_2$ in the plume. Alcohols contribute significantly to the observed signal in the 30--32 u range, while O$_2$ maintains a strong contribution at mass 32 (Fig.~\ref{fig:compare}c). The likelihood that at least one alcohol is present is $> 0.82$, though similarities between their cracking patterns introduce uncertainty in their respective mixing ratios (see Supplementary Information). In their analysis of the low velocity flybys, Waite et al.~\cite{waite_cassini_2017} note that mass 32 exhibits an elevated count rate, likely in part due to surface processing between H$_2$O and the Ti instrument antechamber~\cite{jaramillo-botero_hypervelocity_2012}. Accounting for this instrument effect, they estimated an upper limit on the intrinsic plume O$_2$~$/$~H$_2$O number density of $<0.004\%$. In accounting for additional species, our statistical analysis confirms that the excess at mass channel 32 is best explained by O$_2$, though the overlapping signal from dissociation products of alcohol and CO$_2$ parent molecules precludes the quantification of a precise mixing ratio. That the O$_2$ detected by INMS is in fact native to Enceladus would be consistent with the suite of partially oxidized organics reported here (Fig.~\ref{fig:overview}), and is expected if radiolysis or photolysis of H$_2$O-ice contribute to the observed plume chemistry~\cite{johnson_photolysis_1997}. Nevertheless, further characterization of potential instrument artifacts is warranted to confirm the source of the O$_2$ signal beyond the circumstantial evidence provided here.

\begin{figure}[t]
\centering 
\renewcommand\figurename{Fig.}
\includegraphics[width=\columnwidth]{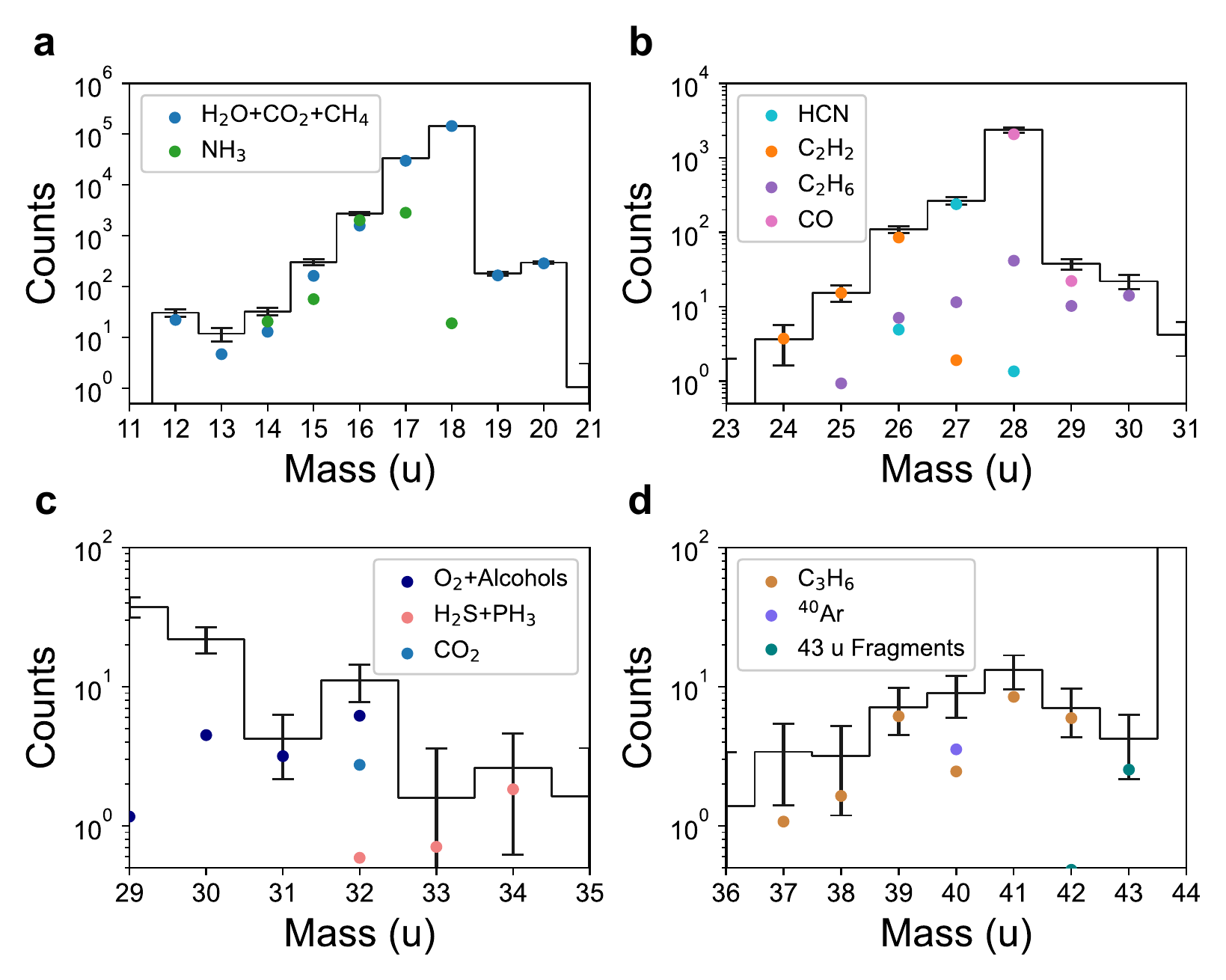}
\caption{\label{fig:compare}Contributions of individual species to the model fit. Black silhouettes show different mass ranges for the average low velocity INMS spectrum plotted in Fig.~\ref{fig:spectrum}. Error bars show $1\sigma$ Gaussian uncertainty in the observed counts. (a) Blue circles show the total contribution of H$_2$O+CO$_2$+CH$_4$. Green circles show the contribution of NH$_3$. (b) Model contributions from HCN, C$_2$H$_2$, C$_2$H$_6$, and CO (cyan, orange, purple, and magenta circles, respectively). (c) Navy circles show the contribution from O$_2$ added to the total contribution from alcohols. The blue circle shows the dissociated O$_2$ contribution from cracked CO$_2$. Pink circles show the tentatively proposed contribution of H$_2$S+PH$_3$. Low signal-to-noise ratios at these mass channels preclude the firm identification of H$_2$S or PH$_3$. (d) Brown and indigo circles show contributions from C$_3$H$_6$ and $^{40}$Ar, respectively. The teal circle shows the total contribution from 43 u fragments.}
\end{figure}

Our analysis also indicates moderate (Table~\ref{tab:mixing}) evidence for either H$_2$S or PH$_3$ in the plume, with probabilities of 0.37 and 0.24, respectively. This conclusion stems primarily from mass 34, where the omission of these species leaves a small residual signal that cannot be explained by other library spectra (Fig.~\ref{fig:compare}c). However, this signal alone is not enough to unambiguously conclude their presence within the INMS data. Both species improve the model fit when added to the minimal model (``Confirmed" species in Table~\ref{tab:mixing}), but not when additional higher likelihood species are included (``Strong" species in Table~\ref{tab:mixing}). Our results are therefore only suggestive of a 34 u compound in the plume. Given the profound astrobiological implications of finding sulfur or phosphorous compounds at Enceladus, mass spectra for alternative 34 u species (such as H$_2$O$_2$) should be characterized experimentally and investigated in future studies.

\section{Discussion}

In aggregate, the results presented here indicate that Enceladus is host to a multiphasic and compositionally diverse chemical environment that is consistent with a habitable subsurface ocean. The new species identified in this work also suggest that this ocean may contain the necessary building blocks required to synthesize compounds important to the origin of life.

The detections of partially oxidized hydrocarbons along with the very likely detections of an alcohol and molecular O$_2$ are particularly interesting as these species potentially imply a diverse redox environment within the ocean (Fig.~\ref{fig:overview}). Past measurements of CH$_4$ and H$_2$ in the plume~\cite{waite_cassini_2006, waite_jr_liquid_2009, waite_cassini_2017} support the hypothesis that Enceladus may be hydrothermally active and could be a source of biologically useful reductants. In particular, methanogenesis via the reduction of CO$_2$ has been proposed as a potential pathway that could support extant microbial communities near the sea floor~\cite{affholder_bayesian_2021, hoehler_implications_2022}. However, without additional oxidants, reductants in the ocean would be of little biochemical utility, as no electron transfer mechanism beyond methanogenesis would be available to yield a negative change in Gibbs free energy. The presence of O$_2$ and partially oxidized carbon compounds may solve this problem, as they implicate a multitude of highly exergonic redox pathways that could help power life in Enceladus' subsurface ocean. Such compounds, for example, could serve as direct substrates for biological growth~\cite{oremland_acetylene_2008, colby_biological_1979, yurimoto_assimilation_2005, singh_anaerobic_2017}, or be intermediaries of other metabolic reactions involving additional organics and oxidants (see e.g., refs.~\cite{russell_methane_2017, zolotov_model_2004, ray_oxidation_2021}).

Our results also provide the first conclusive evidence for HCN, C$_2$H$_2$, and C$_3$H$_6$ in the plume, which, in addition to their significance for habitability, are particularly intriguing for their relevance to prebiotic chemistry. HCN polymerization is implicated in a number of potential pathways for the formation of nucleobases and amino acids~\cite{miller_mechanism_1957, orgel_prebiotic_2004, oro_synthesis_1961, patel_common_2015}. Although these reactions might be limited within a dilute subsurface ocean, concentrated conditions favorable for polymerization could be achieved within the ice shell via eutectic freezing~\cite{miyakawa_cold_2002, orgel_prebiotic_2004, miller_origins_1974}. Indeed, Levy et al.~\cite{levy_prebiotic_2000} demonstrated the production of alanine, glycine, aspartic acid, and adenine from HCN and NH$_3$ under conditions similar to those of icy satellites. Repeated freezing and thawing caused by the cycling of material between the ice shell and the warmer interior fissures of the plume vents may provide conditions favorable for organic synthesis. 

In accordance with this scenario, there does exist evidence that the plume contains at least some material that has not been diluted by the ocean. Liquid water facilitates the rapid hydrolysis of HCN into CH$_3$NO, which subsequently decays into CH$_2$O$_2$ and NH$_3$~\cite{miyakawa_cold_2002}. The simultaneous presence of ppt amounts of HCN and absence of CH$_3$NO and CH$_2$O$_2$ suggests that the plume may be sourced, in part, by solid-phase material residing in the ice shell~\cite{waite_jr_liquid_2009}. HCN at Enceladus could be primordial in nature or produced via the radiolysis of nitrogen-bearing surface ices by magnetospheric electrons. Laboratory experiments of hydrocarbon-rich ices simulating the temperature and surface radiolysis conditions of Enceladus support the evidence for cyano group-containing species. Hand~\cite{hand_physics_2007} and Hand et al.~\cite{hand_laboratory_2006} found that the warming of H$_2$O+NH$_3$+C$_3$H$_6$ ice films from 70 K to 300 K after irradiation by 10 keV electrons produces HCN and nitrile species, including CH$_3$CN. Irradiation of H$_2$O+CO$_2$ ice in the presence of hydrocarbons can also produce alcohols, including CH$_3$OH~\cite{hand_physics_2007}. Thus, the detection of HCN and an alcohol (as well as O$_2$ and possibly CH$_3$CN) in the plume could potentially be explained by the aerosolization of radiolytically processed material on or near the surface. Additional evidence from organic-rich ice grains detected in the plume further suggests the presence of solid-phase organic films that exist above the water table and are transported to the surface via the plume vents~\cite{postberg_macromolecular_2018}. The macromolecular nature of these compounds, some in excess of 200 u, might be evidence of ongoing synthetic chemistry. Accumulation and isolation of buoyant organic compounds at the cold ice-water interface between the ocean and ice shell may also promote longer lifetimes and inhibit hydrolysis~\cite{pappalardo_astrobiology_2009}. 

Of course, the availability of organic compounds at Enceladus to support or facilitate the origin of life likely depends strongly on the geochemistry of the subsurface ocean. Although the mineral composition of the ocean floor is unknown, the simultaneous detections of SiO$_2$ particles by CDA~\cite{hsu_ongoing_2015} and gaseous H$_2$ by INMS~\cite{waite_cassini_2017} indicate the presence of a complex hydrothermal environment~\cite{glein_geochemistry_2018}. Similarities between the inferred temperature and pH of Enceladus’ ocean and those of the Lost City hydrothermal vents point towards serpentinization reactions as a possible source for the observed H$_2$ abundance~\cite{mckay_enceladus_2018, waite_cassini_2017, zolotov_chemical_2011}. Such an explanation would also be consistent with the tentative evidence for H$_2$S presented here. The presence of C$_2$H$_2$ and C$_2$H$_6$ in the plume further implicates ongoing catalytic reactions driven by metal-bearing minerals within the ocean~\cite{matson_enceladus_2007, foustoukos_hydrocarbons_2004, proskurowski_abiogenic_2008}. If ferrous iron is present at Enceladus (as it is at the Lost City sites) along with H$_2$S, sufficient reducing power may be available for a metabolic pathway to abiogenesis~\cite{blochl_reactions_1992, orgel_prebiotic_2004, wachtershauser_before_1988}. Laboratory evidence suggests that confirmation of C$_3$H$_6$ in the plume might allow for the formation of vesicle-type structures at Enceladus, which could then shelter the burgeoning proto-metabolism~\cite{hand_physics_2007}. 

An alternative scenario for complex organic synthesis could be realized through the photochemical processing of ocean material ejected by the plume onto the surface. HCN might be sequestered as ferrocyanide by sodium or potassium salts, both of which have been found in plume ice grains~\cite{postberg_sodium_2009, postberg_salt-water_2011}. Ferrocyanide is an important feedstock molecule for the cyanosulfidic prebiotic chemistry paradigm, which relies on reductive homologation of HCN to form the sugars needed for ribonucleotide assembly~\cite{benner_when_2020, patel_common_2015, sasselov_origin_2020}. If there exists sufficient UV photon processing of organic material on Enceladus’ surface, the selective production of RNA and amino acid precursors from plume-derived HCN and C$_2$H$_2$ (and possibly C$_3$H$_6$O, H$_2$S, and PH$_3$), together with plausible mineralogical catalysts might occur~\cite{patel_common_2015, powner_synthesis_2009}. This surface-based production could then feed back into the plume or ocean via downward transport through the ice shell. Whether this type of chemistry is efficient under Enceladus-like conditions could be explored in future experimental studies, while more detailed examination of Enceladus' oceanic material will require future robotic missions. 

\section{Methods}

\subsection{Instrument overview}

The \textit{Cassini} Ion and Neutral Mass Spectrometer (INMS) was a quadrupole mass spectrometer designed primarily for neutral gas analysis~\cite{waite_cassini_2004}. When INMS was operating in its Closed Source Neutral (CSN) mode, neutral molecules were accumulated in the instrument antechamber before being directed towards a 70 eV electron source for subsequent ionization. Detections of ionized molecules at the instrument target were then recorded sequentially at individual mass channels corresponding to mass-to-charge ratios (m/z), where z is typically assumed to be 1. The INMS mass channels ranged from 1 to 99 u at a resolution of 1 u, excluding 9, 10, and 11 u. 

When neutral molecules enter the ionization region of a mass spectrometer such as INMS, emitted electrons from the electron source both ionize incoming parent molecules and produce dissociated, ionized fragments. For a given electron energy (e.g., 70 eV), each incoming parent molecule fragments according to a specific cracking pattern that describes the relative proportions of the dissociated products. As such, a mass spectrum obtained from a single molecular species will exhibit counts at mass channels pertaining to the molecular weight of the ionized parent molecule, as well as those of each of the ionized dissociation products. INMS spectra therefore consist of a combination of overlapping spectral features resulting from each parent molecule’s cracking pattern. 

INMS was not able to detect parent molecules or fragments with masses above the maximum mass of 99 u. However, larger molecules may have impacted the instrument walls and fragmented into smaller molecules that were within the detectable mass range~\cite{waite_jr_liquid_2009}. The rate of these impact fragmentation events depends heavily on the kinetic energy of the spacecraft~\cite{postberg_plume_2018}. As such, spectra generated from faster flybys ($>$ 10--15 km/s) exhibit a suite of spectral features (associated with impact fragments of higher mass molecules) that are not observed at lower speeds~\cite{postberg_macromolecular_2018, postberg_plume_2018, waite_cassini_2017}. As a result, measurements obtained during the lower velocity flybys of Enceladus provide the best opportunity to study the intrinsic plume composition in the absence of fragmentation effects. 

\subsection{Data set selection}

During the E14, E17, and E18 flybys of Enceladus, \textit{Cassini} made a series of three low velocity ($< 7.5$ km/s) passages through the plume along nearly identical trajectories. These flybys, referred to as the “slow” flybys~\cite{waite_cassini_2017}, produced the most consistent INMS data with which to analyze the plume. Previous studies~\cite{waite_cassini_2017, waite_jr_liquid_2009, waite_cassini_2006, postberg_macromolecular_2018, magee_inms-derived_2009} have developed a robust data processing procedure for converting the raw time series data collected by INMS into instrument-corrected mass spectra. The post-processed slow flyby INMS spectra have been used in other compositional analyses~\cite{waite_cassini_2017, postberg_macromolecular_2018} and provide the best opportunity to study the plume’s intrinsic neutral gas composition in the absence of instrument artifacts. Here, we use the averaged CSN spectrum obtained across the three slow flybys, as done in refs.~\cite{waite_cassini_2017, postberg_macromolecular_2018}. This averaged slow flyby spectrum is identical to that presented in ref.~\cite{postberg_macromolecular_2018}, with minor differences described below.

In order to quantify variability in the INMS data between flybys, we assumed a Gaussian random error on the counts at each mass channel. The $2\sigma$ width for each count distribution was estimated based on the range of observed counts at each mass channel across the three flybys (as depicted in ref.~\cite{postberg_plume_2018}), similar to that described by Waite et al. in their own analysis of the slow flyby spectrum~\cite{waite_cassini_2017}. However, mass 18 is the anchor mass used to standardize the E14, E17, and E18 flybys and therefore, by definition, does not exhibit any inter-flyby variability. Other H$_2$O-dominated mass channels at 17, 19, and 20 u also demonstrate reduced variability as a result of this standardization. To account for this effect, we instead used a standard Poisson uncertainty equal to the square root of the average number of counts on all mass channels where the inter-flyby variability is smaller than the uncertainty due to counting statistics (note that in the limit of large counts, the Poisson distribution approaches the Gaussian distribution). Moreover, mass channels beyond 46 u exhibit significantly lower signal-to-noise ratios than lower mass channels. To avoid erroneously fitting to these data points, we implemented a larger fractional uncertainty on peaks where both the inter-flyby variability and the counting statistics error is less than the typical count value measured at these mass channels ($\sim$2 counts). This larger relative uncertainty was set equal to the magnitude of this noisy signal and serves as the minimum uncertainty across all mass channels.

When INMS was operating in the CSN mode, the vast majority of counts at 1 and 2 u arose due to interactions between gas phase H$_2$O molecules and the walls of the instrument antechamber~\cite{waite_cassini_2017}. Proper assessment of these mass channels requires a careful analysis of data from the instrument’s Open Source Neutral Beam (OSNB) mode. This problem has been explored in detail by Waite et al.~\cite{waite_cassini_2017} who attribute $\sim$98\% of the total signal at 1 and 2 u to H$_2$ (at a mixing ratio of $\sim$0.9\%) and H$_2$O. The cracking pattern for H$_2$ does not contribute to any other mass channels and is therefore independent from the deconvolution of the remainder of the spectrum. Furthermore, the mixing ratio of H$_2$O is predominantly dictated by high leverage points at masses 18 and 17 and not significantly affected by these lower mass channels. For these reasons, we chose to adopt the H$_2$ mixing ratio of ref.~\cite{waite_cassini_2017} and exclude masses 1 and 2 from our analysis (Extended Data Fig.~\ref{fig:full_spectrum}).

\subsection{Spectral library}

As discussed in the main text, species that are present within the INMS spectrum cannot be identified unless explicitly included within a reference library used for comparison. However, large spectral libraries suffer from high dimensionality; the inclusion of too many library species leads to models that are unnecessarily complex and prone to overfitting. This phenomenon is similar to how any univariate relationship can be fit with zero residual error using a sufficiently high order polynomial. Although overfitting can be ameliorated by using regularizing heuristics such as the AICc, chemically implausible species may still be incorporated into the final model if they are included within the initial library. As an additional hindrance, the inclusion of many (possibly collinear) model parameters reduces statistical power and raises the computational intensity of the analysis. It is therefore desirable to constrain the library of candidate species as much as possible based on prior knowledge and the results of previous studies.

Our library consists of a carefully chosen set of 50 compounds, including the most recent list of INMS-detectable candidate plume species presented in ref.~\cite{postberg_plume_2018}. All library species and their reasons for inclusion are presented in Extended Data Table~\ref{tab:library}. Because all counts observed above 46 u during the slow flybys are below the estimated minimum uncertainty, no species with base peaks above this threshold were included in this analysis. The spectra in our library are 70 eV electron ionization mass spectra collected from the INMS Refurbished Engineering Unit (REU) and the National Institute of Standards and Technology (NIST) online database. We adopted the method used in refs.~\cite{magee_inms-derived_2009, miller_cassini_2020} to prioritize the REU spectra when available. REU spectra were obtained from the most recent INMS calibration file provided on the Planetary Data System. All spectra were analyzed at 1 u resolution and matched to the mass detection range of INMS. All spectra were base peak normalized to ensure a uniform method of comparison.

\subsection{Benchmarking model complexity}

In deducing limits on INMS model complexity, we constructed all possible combinations of plume species up to $d=15$ from a candidate library of 50 plausible compounds (Extended Data Table~\ref{tab:library}). To reduce the computational intensity associated with this analysis, we assumed H$_2$O, CO$_2$, and CH$_4$ to be present in each model. H$_2$O and CO$_2$ have been verified at Enceladus by independent Cassini instruments~\cite{hansen_enceladus_2006, brown_composition_2006}, and CH$_4$ is among the most consistent INMS observations, having been detected on every flyby that sampled the plume~\cite{waite_cassini_2006, waite_jr_liquid_2009, waite_cassini_2017}. This resulted in a total of $\sum_{d=3}^{15} \frac{47!}{(d-3)!(50-d)!} \approx 7.7 \times 10^{10}$ models for direct comparison. In order to efficiently sample the parameter space of more complex models, we implemented a forward modeling stepwise selection procedure~\cite{james_introduction_2013}. Starting from the three-component model of H$_2$O$+$CO$_2$$+$CH$_4$, the next most complex model was constructed by including an additional species that, when added to the model, resulted in the greatest decrease in the variance-weighted sum of squared residuals. This selection process was repeated—sequentially incorporating new species based on their influence on the sum of squared residuals—to produce a set of nested models, each containing one more species than the last. Notably, the sum of squared residuals (i.e., the training error) was only used to compare models of equal complexity and therefore selects the same species as would comparisons based on the AICc. As shown in Fig.~\ref{fig:complex}, the results of this forward modeling algorithm resemble those of the exhaustive model search for $d < 16$. Extrapolation of these results to more complex models confirms a monotonic decrease in relative model likelihood for $d > 13$.

\subsection{Multi-model inference}

Initial consideration of the exhaustive set of possible models reduces user bias and ensures that all plausible compounds are accounted for. Such a complete analysis also allows for a robust determination of optimal model complexity (Fig.~\ref{fig:complex}). However, sound statistical inferences cannot---and should not---be made using the exhaustive model list. The inclusion of extraneous, low likelihood models into the final inferential set precludes reliable parameter estimation due to the well-known bias-variance tradeoff phenomenon (for a complete discussion see, e.g., ref.~\cite{burnham_model_2004}). Therefore, in order to extract a reduced set of high likelihood models suitable for multi-model inference, we eliminated all models with $\lambda < 1/e^3$, as they are exponentially less likely to explain the INMS data. Monte Carlo studies have demonstrated that this threshold corresponds to a $\sim$95--99\% model confidence interval, leading to its broad acceptance within the literature~\cite{burnham_model_2004, richards_dealing_2007, richards_model_2011, richards_testing_2005}. An extensive post-hoc analysis further suggests that our results are stable to alternative model selection criteria (see Supplementary Information).

In order to determine mixing ratios that properly account for the ambiguity induced by model degeneracy, we computed model-averaged regression coefficients based on the relative likelihood of each model,
\begin{equation}\label{eq:average}
    \bar{\beta}_k = \sum^R_{j=1} w_j \beta_{k,j}.
\end{equation}
Here, $\beta_{k,j}$ is the regression coefficient for species $k$ in model $j$, $R$ is the total number of most probable models (those with $\lambda > 1/e^3$), and $w_j = \lambda_j / \sum_m^R \lambda_m$ is the Akaike weight~\cite{burnham_model_2004} of each model. Because the observed INMS spectrum consists of sample data taken from the unknown population distribution of plume contents, the minimum AICc model is a statistical random variable that estimates the actual (also unknown) model that minimizes the Kullback-Leibler divergence with the true distribution. Therefore, the Akaike weight may be interpreted as the probability that a given model minimizes this information loss between the model and the unknown, true distribution from which the observed INMS data were sampled. The probability that each species is present in the true minimum AICc model follows accordingly as,
\begin{equation}
    \mathcal{P}_k = \sum^R_{j=1} w_j \Theta(\beta_{k,j})
\end{equation}
where $\Theta(\beta_{k,j})$ is the Heaviside step-function that equals 1 when $\beta_{k,j} > 0$ and is zero otherwise. We then computed uncertainties in the model parameters using the unconditional standard error estimator~\cite{burnham_model_2004},
\begin{equation}
    \mathrm{SE}(\bar{\beta}_k) = \sum^R_{j=1} w_j \sqrt{\mathrm{var}(\beta_{k,j}) + (\beta_{k,j} - \bar{\beta}_k)^2}
\end{equation}
where $\mathrm{var}(\beta_{k,j})$ is the variance of the regression coefficient conditional on model $M_j$. Here, $\mathrm{var}(\beta_{k,j})$ characterizes the intra-model uncertainty associated with maximum likelihood estimation, whereas the term $(\beta_{k,j} - \bar{\beta}_k)^2$ quantifies inter-model variability due to the presence of additional high likelihood models. Importantly, the results presented in this work incorporate the relative probabilities of each model and account for ambiguities in the model selection process.

\subsection{Conversion to mixing ratios}

The formula for converting INMS counts to ambient densities is described in ref.~\cite{teolis_revised_2015} and given by,
\begin{equation}
    n_k = \left(\frac{T_0}{T_a}\right)^{1/2} \frac{1}{D_k} \frac{X_k}{s_k}
\end{equation}
where $X_k$ is the count rate of species $k$ (measured at the base peak), $s_k$ is the INMS sensitivity, $D_k$ is the ram enhancement factor, $T_a$ is the ambient temperature, and $T_0$ is room temperature (293 K). When available, INMS sensitivities were obtained from refs.~\cite{miller_cassini_2020, magee_inms-derived_2009, cui_analysis_2009} and otherwise estimated based on the electron impact ionization cross section procedure of Fitch and Sauter (see Eq. 2 of ref.~\cite{fitch_calculation_1983}) adapted to INMS data in refs.~\cite{miller_cassini_2020, magee_inms-derived_2009}. For one species (PH$_3$), cross section data was taken from ref.~\cite{graves_calculated_2021}. As done in ref.~\cite{miller_cassini_2020}, a 30\% uncertainty was implemented on all sensitivities estimated from NIST spectra.

The count rate is related to the model-averaged regression coefficient of Equation \eqref{eq:average} by $X_k= \bar{\beta}_k c_0$ where $c_0$ is the standardized (species-independent) base peak count rate of each library spectrum. The mixing ratio for each species $m_k$ (scaled to include the H$_2$ mixing ratio $m_{\text{H}_2} = 0.9\%$ given in ref.~\cite{waite_cassini_2017}) is then,
\begin{align}
    m_k &= (100 - m_{\text{H}_2}) \frac{n_k}{\sum_l n_l} \nonumber\\
    &= (100 - m_{\text{H}_2}) \frac{\bar{\beta}_k}{D_k s_k} \sum_l \frac{D_l s_l}{\bar{\beta}_l}.
\end{align}
At suprathermal spacecraft speed and small ram angle (conditions valid during the E14, E17, and E18 flybys), the ram enhancement factor is approximately~\cite{waite_cassini_2017, teolis_revised_2015},
\begin{equation}
    D_k \sim 0.7 u \sqrt{\frac{2\pi \mu_k}{k_B T_a}}
\end{equation}
for spacecraft speed $u$ and molecular mass $\mu_k$. The final expression for the mixing ratio of each species is therefore,
\begin{equation}
    m_k = (100 - m_{\text{H}_2}) \frac{\bar{\beta}_k}{\sqrt{\mu_k} s_k} \sum_l \frac{\sqrt{\mu_l} s_l}{\bar{\beta}_l}.
\end{equation}
Future updates to INMS sensitivity coefficients or ram enhancement factors may change the relative mixing ratios presented here but will not affect which species have been identified---or the evidence for their detection---as these depend only on the set of $\{\bar{\beta}_k\}$.

\section{Data availability}
All INMS REU spectra are publicly available from the Planetary Data System at \href{https://doi.org/10.17189/1519605}{https://doi.org/10.17189/1519605}. All NIST spectra are publicly available from the NIST Chemistry WebBook at \href{https://doi.org/10.18434/T4D303
}{https://doi.org/10.18434/T4D303}.

\section{Code availability}
The code for this work was implemented using open source Python libraries available at \href{https://pypi.org/}{https://pypi.org/} and is available upon request.

\section{Additional Information}

Correspondence and requests for materials should be addressed to J.S.P.

\section{Acknowledgements}

The authors thank J. Hunter Waite and Brian A. Magee for their help on interpreting INMS instrument effects. J.S.P. thanks Masao Sako and Kiri L. Wagstaff for useful discussions and statistical insight, as well as Dimitar D. Sasselov for helpful discussions on prebiotic chemistry. All authors acknowledge the support of the Cassini Data Analysis Program (NNN13D466T) and the Jet Propulsion Laboratory, California Institute of Technology, under a contract with NASA. T.A.N. was also supported by an appointment to the NASA Postdoctoral Fellowship Program at the Jet Propulsion Laboratory administered by Oak Ridge Associated Universities and Universities Space Research Association through a contract with NASA. K.P.H. also acknowledges support from the NASA Astrobiology Program (80NSSC19K1427) and the Europa Lander Pre-Project, managed by the Jet Propulsion Laboratory, California Institute of Technology, under a contract with NASA.

\section{Author contributions}

J.S.P, T.A.N., and K.P.H contributed to the conceptual development of the study, the interpretation of results, and to the final version of the manuscript. J.S.P. developed the methodology, performed the mathematical analysis, and wrote the first version of the manuscript. 

\section{Competing interests}

The authors declare no competing interests.

\begin{table*}[p]
\centering
\begin{threeparttable}
\renewcommand\tablename{Extended Data Table}
\setcounter{table}{0}
\caption{\label{tab:library}Complete list of species included in the analysis. Those taken from ref.~\cite{postberg_plume_2018} comprise the most recently published list of INMS-detectable species in the plume. REU: INMS Refurbished Engineering Unit; NIST: National Institute of Standards and Technology online database.}
\begin{tabular*}{\textwidth}{lllllll}
\toprule
Species & Name & Mass (u) & Source & $s_k$ & $s_k$ Ref. & Reason/Ref.\\
\midrule
CH$_4$	&	Methane	&	16	&	REU	&	6.01$\times 10^{4}$  &~\cite{miller_cassini_2020}	&	\cite{postberg_plume_2018, waite_cassini_2017}	\\
NH$_3$	&	Ammonia	&	17	&	REU	&	4.77$\times 10^{4}$  &~\cite{miller_cassini_2020}	&	\cite{postberg_plume_2018, waite_cassini_2017}	\\
H$_2$O	&	Water	&	18	&	REU	&	4.34$\times 10^{4}$  &~\cite{miller_cassini_2020}	&	\cite{postberg_plume_2018, waite_cassini_2017}	\\
C$_2$H$_2$	&	Acetylene	&	26	&	REU	&	8.81$\times 10^{4}$  &~\cite{miller_cassini_2020}	&	\cite{postberg_plume_2018, waite_cassini_2006}	\\
HCN	&	Hydrogen Cyanide	&	27	&	REU	&	5.20$\times 10^{4}$  &~\cite{miller_cassini_2020}	&	\cite{postberg_plume_2018, waite_jr_liquid_2009}	\\
C$_2$H$_4$	&	Ethylene	&	28	&	REU	&	6.21$\times 10^{4}$  &~\cite{miller_cassini_2020}	&	\cite{postberg_plume_2018, waite_cassini_2017}	\\
CO	&	Carbon Monoxide	&	28	&	REU	&	6.60$\times 10^{4}$  &~\cite{miller_cassini_2020}	&	\cite{postberg_plume_2018, waite_cassini_2017}	\\
N$_2$	&	Nitrogen	&	28	&	REU	&	6.29$\times 10^{4}$  &~\cite{miller_cassini_2020}	&	\cite{postberg_plume_2018, waite_cassini_2017}	\\
CH$_2$O	&	Formaldehyde	&	30	&	NIST	&	3.21$\times 10^{4}$  &~\cite{miller_cassini_2020}	&	\cite{postberg_plume_2018, waite_jr_liquid_2009}	\\
NO	&	Nitric Oxide	&	30	&	NIST	&	5.28$\times 10^{4}$  & $-$	&	\cite{postberg_plume_2018}	\\
C$_2$H$_6$	&	Ethane	&	30	&	REU	&	6.94$\times 10^{4}$  &~\cite{miller_cassini_2020}	&	\cite{postberg_plume_2018, waite_jr_liquid_2009}	\\
CH$_5$N	&	Methylamine	&	31	&	NIST	&	4.07$\times 10^{4}$  &~\cite{miller_cassini_2020}	&	\cite{postberg_plume_2018}	\\
CH$_3$OH	&	Methanol	&	32	&	NIST	&	4.17$\times 10^{4}$  &~\cite{miller_cassini_2020}	&	\cite{postberg_plume_2018, waite_cassini_2017}	\\
H$_2$S	&	Hydrogen Sulfide	&	34	&	NIST	&	5.38$\times 10^{4}$  &~\cite{miller_cassini_2020}	&	\cite{postberg_plume_2018, waite_jr_liquid_2009}	\\
PH$_3$	&	Phospine	&	34	&	NIST	&	5.98$\times 10^{4}$  & $-$	&	\cite{postberg_plume_2018}	\\
O$_2$	&	Oxygen	&	36	&	NIST	&	5.03$\times 10^{4}$  &~\cite{miller_cassini_2020}	&	\cite{postberg_plume_2018, waite_cassini_2017}	\\
$^{36}$Ar	&	Argon 36	&	36	&	REU	&	7.87$\times 10^{4}$  &~\cite{cui_analysis_2009}	&	\cite{postberg_plume_2018, waite_cassini_2017}	\\
C$_3$H$_4$	&	Propyne	&	40	&	REU	&	4.19$\times 10^{4}$  &~\cite{miller_cassini_2020}	&	\cite{postberg_plume_2018, waite_jr_liquid_2009}	\\
$^{40}$Ar	&	Argon 40	&	40	&	REU	&	7.29$\times 10^{4}$  &~\cite{magee_inms-derived_2009}	&	\cite{postberg_plume_2018, waite_jr_liquid_2009}	\\
CH$_3$CN	&	Acetonitrile	&	41	&	REU	&	5.28$\times 10^{4}$  &~\cite{miller_cassini_2020}	&	\cite{postberg_plume_2018, hand_physics_2007}	\\
C$_2$H$_2$O	&	Ketene	&	42	&	NIST	&	4.37$\times 10^{4}$  &~\cite{miller_cassini_2020}	&	\cite{postberg_plume_2018}	\\
C$_3$H$_6$	&	Propene	&	42	&	NIST	&	4.06$\times 10^{4}$  &~\cite{miller_cassini_2020}	&	\cite{postberg_plume_2018, waite_jr_liquid_2009}	\\
CH$_2$N$_2$	&	Cyanamide	&	42	&	NIST	&	7.55$\times 10^{4}$  & $-$	&	Organic synthesis~\cite{powner_synthesis_2009, benner_when_2020}	\\
C$_2$H$_4$O	&	Acetaldehyde	&	44	&	NIST	&	4.42$\times 10^{4}$  &~\cite{miller_cassini_2020}	&	\cite{postberg_plume_2018, waite_jr_liquid_2009}	\\
C$_3$H$_8$	&	Propane	&	44	&	REU	&	5.11$\times 10^{4}$  &~\cite{miller_cassini_2020}	&	\cite{postberg_plume_2018, waite_cassini_2006}	\\
CO$_2$	&	Carbon Dioxide	&	44	&	REU	&	7.09$\times 10^{4}$  &~\cite{miller_cassini_2020}	&	\cite{postberg_plume_2018, waite_cassini_2017}	\\
C$_2$H$_7$N	&	Ethylamine	&	45	&	NIST	&	1.71$\times 10^{4}$  &~\cite{miller_cassini_2020}	&	\cite{postberg_plume_2018}	\\
CH$_3$NO	&	Formamide	&	45	&	NIST	&	6.22$\times 10^{4}$  &~\cite{miller_cassini_2020}	&	HCN decomposition~\cite{miyakawa_cold_2002}	\\
C$_2$H$_6$O	&	Ethanol	&	46	&	NIST	&	1.62$\times 10^{4}$  &~\cite{miller_cassini_2020}	&	\cite{postberg_plume_2018, waite_jr_liquid_2009}	\\
CH$_2$O$_2$	&	Formic Acid	&	46	&	NIST	&	2.94$\times 10^{4}$  &~\cite{miller_cassini_2020}	&	HCN decomposition~\cite{miyakawa_cold_2002}	\\
C$_4$H$_8$	&	1-Butene	&	56	&	NIST	&	3.73$\times 10^{4}$  &~\cite{miller_cassini_2020}	&	\cite{postberg_plume_2018, waite_jr_liquid_2009}	\\
C$_2$H$_6$N$_2$	&	Azomethane	&	58	&	NIST	&	8.46$\times 10^{4}$  & $-$	&	\cite{postberg_plume_2018}	\\
C$_3$H$_6$O	&	Acetone	&	58	&	NIST	&	6.70$\times 10^{4}$  &~\cite{miller_cassini_2020}	&	\cite{postberg_plume_2018, waite_jr_liquid_2009}	\\
C$_4$H$_{10}$	&	Isobutane	&	58	&	NIST	&	3.24$\times 10^{5}$  &~\cite{miller_cassini_2020}	&	\cite{postberg_plume_2018, waite_jr_liquid_2009}	\\
C$_2$H$_4$O$_2$	&	Acetic Acid	&	60	&	NIST	&	4.51$\times 10^{4}$  &~\cite{miller_cassini_2020}	&	\cite{postberg_plume_2018, waite_jr_liquid_2009}	\\
C$_3$H$_8$O	&	1-Propanol	&	60	&	NIST	&	8.80$\times 10^{5}$  &~\cite{miller_cassini_2020}	&	\cite{postberg_plume_2018, waite_jr_liquid_2009}	\\
C$_2$H$_7$NO	&	Monoethanolamine	&	61	&	NIST	&	1.67$\times 10^{3}$  & $-$	&	\cite{postberg_plume_2018}	\\
C$_2$H$_6$O$_2$	&	1,2-Ethanediol	&	62	&	NIST	&	5.77$\times 10^{5}$  &~\cite{miller_cassini_2020}	&	\cite{postberg_plume_2018, waite_jr_liquid_2009}	\\
C$_5$H$_{10}$	&	Cyclopentane	&	70	&	NIST	&	4.39$\times 10^{4}$  &~\cite{miller_cassini_2020}	&	\cite{postberg_plume_2018}	\\
C$_4$H$_9$N	&	Pyrrolidine	&	71	&	NIST	&	1.18$\times 10^{3}$  & $-$	&	\cite{postberg_plume_2018}	\\
C$_5$H$_{12}$	&	Pentane	&	72	&	NIST	&	1.53$\times 10^{4}$  &~\cite{miller_cassini_2020}	&	\cite{postberg_plume_2018, waite_jr_liquid_2009}	\\
C$_4$H$_{10}$O	&	1-Butanol	&	74	&	NIST	&	6.14$\times 10^{5}$  & $-$	&	\cite{postberg_plume_2018}	\\
C$_2$H$_5$NO$_2$	&	Glycine	&	75	&	NIST	&	6.14$\times 10^{5}$  &~\cite{miller_cassini_2020}	&	\cite{postberg_plume_2018, waite_jr_liquid_2009}	\\
C$_3$H$_5$Cl	&	Allyl Chloride	&	76	&	NIST	&	9.63$\times 10^{4}$  & $-$	&	\cite{postberg_plume_2018}	\\
C$_5$H$_9$N	&	Butyl Isocyanide	&	83	&	NIST	&	8.34$\times 10^{4}$  & $-$	&	Hydrocarbon irradiation~\cite{hand_physics_2007}	\\
C$_4$H$_6$O$_2$	&	2,3-Butanedione	&	86	&	NIST	&	3.91$\times 10^{4}$  &~\cite{miller_cassini_2020}	&	\cite{postberg_plume_2018, waite_jr_liquid_2009}	\\
C$_3$H$_7$NO$_2$	&	Alanine	&	89	&	NIST	&	1.73$\times 10^{3}$  & $-$	&	\cite{postberg_plume_2018}	\\
C$_8$H$_{18}$	&	Octane	&	114	&	NIST	&	1.65$\times 10^{3}$  & $-$	&	\cite{postberg_plume_2018}	\\
C$_6$H$_{12}$N$_4$	&	Methenamine	&	140	&	NIST	&	3.86$\times 10^{3}$  & $-$	&	\cite{postberg_plume_2018}	\\
C$_6$H$_{14}$N$_2$O$_2$	&	Lysine	&	146	&	NIST	&	1.34$\times 10^{3}$  & $-$	&	Amino acid~\cite{glavin_search_2020}	\\
\bottomrule
\end{tabular*}
\end{threeparttable}
\end{table*}

\begin{table*}[p]
\centering
\begin{threeparttable}
\renewcommand\tablename{Extended Data Table}
\caption{\label{tab:full_data}Modeling results for the complete spectral library. Mixing ratios are mean $\pm$ SE and scaled to include the 0.9\% H$_2$ abundance of ref.~\cite{waite_cassini_2017} unless otherwise specified. For compounds with upper limits (mean + 3 SE) less than the minimum INMS count uncertainty (2 counts), upper limits were instead estimated relative to H$_2$O based on a 2 count signal after correcting for ram enhancement and instrument sensitivities. The quantities $N_d$ indicate the number of high likelihood ($\lambda > 1/e^3$) models containing each species for models of dimension $d$. Values in parentheses denote the total number of high likelihood models for each $d$.}
\begin{tabular*}{\textwidth}{lllllllll}
\toprule
Species  & Probability & Mixing Ratio (\%) & $N_{10}$ (4) & $N_{11}$ (146) & $N_{12}$ (1348) & $N_{13}$ (5956) & $N_{14}$ (11134) & $N_{15}$ (7744) \\
\midrule
CH$_4$               	& 1    & $0.11 \pm 0.02$                  & 4 & 146 & 1348 & 5956 & 11134 & 7744 \\
NH$_3$	                & 1    & $1.8 \pm 0.1$                    & 4 & 146 & 1348 & 5956 & 11134 & 7744 \\
H$_2$O	                & 1    & $95.9 \pm 0.3$                   & 4 & 146 & 1348 & 5956 & 11134 & 7744 \\
C$_2$H$_2$           	& 0.92 & $0.023 \pm 0.005$                & 3 & 110 & 1103 & 5098 & 10048 & 7646 \\
HCN	                    & 1    & $0.11 \pm 0.02$                  & 4 & 146 & 1348 & 5956 & 11134 & 7744 \\
C$_2$H$_4$	            & 0.10 & $< 0.05$                         & 1 & 36 & 260 & 965 & 1326 & 323 \\
CO	                    & 1    & $0.72 \pm 0.07$                  & 4 & 146 & 1348 & 5956 & 11134 & 7744 \\
N$_2$               	& 0.02 & $< 0.005$                        & 0 & 0 & 15 & 114 & 255 & 255 \\
CH$_2$O                	& 0.07 & $< 0.004$                        & 0 & 2 & 73 & 446 & 921 & 666 \\
NO	                    & 0.05 & $< 0.002$                        & 0 & 7 & 99 & 349 & 695 & 340 \\
C$_2$H$_6$	            & 0.70 & $0.013 \pm 0.010$                & 4 & 102 & 838 & 3850 & 7484 & 5757 \\
CH$_5$N	                & 0.14 & $< 0.003$                        & 0 & 6 & 102 & 647 & 1673 & 1329 \\
CH$_3$OH                & 0.44 & $< 0.005$                        & 2 & 66 & 691 & 2844 & 4634 & 3285 \\
H$_2$S	                & 0.37 & $< 0.003$                        & 0 & 16 & 226 & 1396 & 3701 & 4362 \\
PH$_3$               	& 0.24 & $< 0.002$                        & 0 & 12 & 147 & 963 & 2825 & 2493 \\
O$_2$	                & 0.64 & $< 0.008$                        & 1 & 42 & 511 & 3206 & 7445 & 5575 \\
$^{36}$Ar	            & 0.03 & $< 4 \times 10^{-4}$\tnote{a}    & 0 & 1 & 21 & 155 & 370 & 396 \\
C$_3$H$_4$	            & 0.16 & $< 0.002$                        & 0 & 3 & 127 & 919 & 2416 & 1000 \\
$^{40}$Ar	            & 0.58 & $< 0.004$                        & 0 & 12 & 257 & 2021 & 6293 & 6443 \\
CH$_3$CN	            & 0.16 & $< 0.003$                        & 0 & 5 & 133 & 931 & 2226 & 1234 \\
C$_2$H$_2$O	            & 0.02 & $< 7 \times 10^{-4}$\tnote{a}    & 0 & 0 & 15 & 116 & 260 & 226 \\
C$_3$H$_6$	            & 0.87 & $0.004 \pm 0.002$              & 4 & 143 & 1258 & 5293 & 9278 & 6461 \\
CH$_2$N$_2$	            & 0.06 & $5 \times 10^{-4}$               & 0 & 0 & 23 & 250 & 780 & 801 \\
C$_2$H$_4$O	            & 0.08 & $< 0.002$                        & 0 & 11 & 119 & 574 & 1056 & 442 \\
C$_3$H$_8$	            & 0.23 & $< 0.005$                        & 0 & 21 & 295 & 1332 & 2684 & 1669 \\
CO$_2$	                & 1    & $0.45 \pm 0.04$                  & 4 & 146 & 1348 & 5956 & 11134 & 7744 \\
C$_2$H$_7$N         	& 0.02 & $< 0.002$\tnote{a}               & 0 & 0 & 15 & 144 & 255 & 255 \\
CH$_3$NO	            & 0.02 & $< 5 \times 10^{-4}$\tnote{a}    & 0 & 0 & 15 & 114 & 255 & 255 \\
C$_2$H$_6$O	            & 0.02 & $< 0.002$\tnote{a}               & 0 & 0 & 18 & 127 & 268 & 227 \\
CH$_2$O$_2$	            & 0.02 & $< 0.001$\tnote{a}               & 0 & 0 & 15 & 144 & 255 & 255 \\
C$_4$H$_8$	            & 0.03 & $< 7 \times 10^{-4}$\tnote{a}    & 0 & 1 & 15 & 130 & 330 & 313 \\
C$_2$H$_6$N$_2$      	& 0.24 & $< 0.002$                        & 0 & 15 & 167 & 978 & 2295 & 2587 \\
C$_3$H$_6$O	            & 0.17 & $< 0.001$                        & 0 & 14 & 141 & 748 & 1739 & 1707 \\
C$_4$H$_{10}$	        & 0.08 & $< 0.01$                         & 0 & 7 & 99 & 520 & 933 & 689 \\
C$_2$H$_4$O$_2$	        & 0.02 & $< 6 \times 10^{-4}$\tnote{a}    & 0 & 1 & 16 & 116 & 261 & 225 \\
C$_3$H$_8$O	            & 0.11 & $< 0.005$                        & 0 & 5 & 74 & 481 & 1347 & 1132 \\
C$_2$H$_7$NO	        & 0.23 & $< 0.002$                        & 0 & 25 & 299 & 1432 & 2744 & 1799 \\
C$_2$H$_6$O$_2$     	& 0.20 & $< 0.01$                         & 1 & 19 & 156 & 944 & 2203 & 1899 \\
C$_5$H$_{10}$       	& 0.03 & $< 6 \times 10^{-4}$\tnote{a}    & 0 & 0 & 17 & 155 & 377 & 323 \\
C$_4$H$_9$N	            & 0.06 & $< 2 \times 10^{-4}$\tnote{a}    & 0 & 7 & 83 & 415 & 732 & 565 \\
C$_5$H$_{12}$	        & 0.07 & $< 0.002$\tnote{a}               & 0 & 4 & 66 & 410 & 897 & 553 \\
C$_4$H$_{10}$O	        & 0.03 & $< 4 \times 10^{-4}$\tnote{a}    & 0 & 3 & 36 & 203 & 403 & 305 \\
C$_2$H$_5$NO$_2$    	& 0.08 & $< 0.02$                         & 0 & 14 & 148 & 603 & 911 & 553 \\
C$_3$H$_5$Cl	        & 0.23 & $< 0.001$                        & 0 & 2 & 103 & 845 & 2749 & 2789 \\
C$_5$H$_9$N	            & 0.05 & $< 3 \times 10^{-4}$\tnote{a}    & 0 & 4 & 58 & 313 & 666 & 418 \\
C$_4$H$_6$O$_2$	        & 0.08 & $< 6 \times 10^{-4}$\tnote{a}    & 0 & 7 & 100 & 545 & 902 & 607 \\
C$_3$H$_7$NO$_2$	    & 0.05 & $< 5 \times 10^{-4}$             & 0 & 2 & 41 & 267 & 668 & 387 \\
C$_8$H$_{18}$	        & 0.05 & $< 1 \times 10^{-4}$\tnote{a}    & 0 & 4 & 58 & 315 & 591 & 374 \\
C$_6$H$_{12}$N$_4$   	& 0.05 & $< 5 \times 10^{-5}$\tnote{a}    & 0 & 0 & 20 & 218 & 660 & 670 \\
C$_6$H$_{14}$N$_2$O$_2$	& 0.02 & $< 1 \times 10^{-4}$\tnote{a}    & 0 & 1 & 15 & 116 & 261 & 226 \\
\bottomrule
\end{tabular*}
\begin{tablenotes}
\item[a]{Denotes number density relative to H$_2$O.}
\end{tablenotes}
\end{threeparttable}
\end{table*}

\begin{figure*}[p]
\centering 
\includegraphics[width=\textwidth]{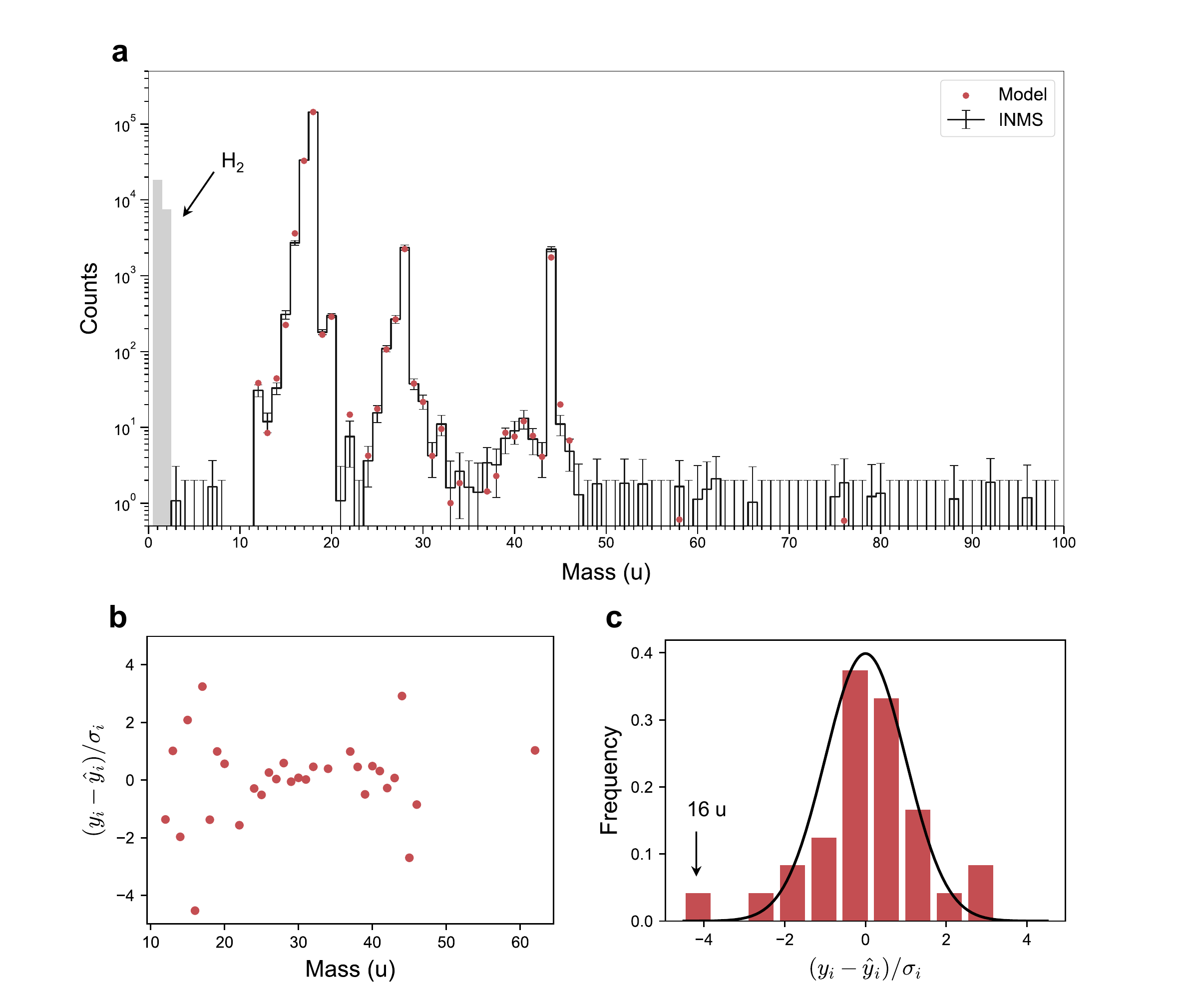}
\renewcommand\figurename{Extended Data Fig.}
\setcounter{figure}{0}
\caption{\label{fig:full_spectrum}Full mass range INMS spectrum and analysis of residuals. (a) The black silhouette shows the full mass range of the INMS spectrum used in this work. Shaded gray bars indicate count values at 1 and 2 u that were omitted from this work (see Methods). The spectrum was adapted from ref.~\cite{postberg_macromolecular_2018}. Error bars show 1$\sigma$ Gaussian uncertainty in the observed count rates. The minimum count uncertainty was estimated as $\sim$2 counts from the count rates of noisy mass channels at masses $> 46$ u. Red circles show the reconstructed model fit based on the mixing ratios of Extended Data Table~\ref{tab:full_data}. (b) Scatterplot of the standardized residuals produced by fitting the model. Only mass channels with counts above the minimum uncertainty are shown. There is no discernable pattern amongst the residuals or evidence of heteroscedasticity. (c) Histogram of the standardized residuals (red bars) compared to a reference Gaussian distribution with zero mean (black curve). The residuals show good agreement with the Gaussian distribution, indicating a robust model fit. The black arrow indicates a potential outlier at mass 16, which likely results from the standardization of the slow flyby count rates at neighboring mass channels.}
\end{figure*}

\end{document}


\title{Detection of HCN and diverse redox chemistry in the plume of Enceladus:\\
SUPPLEMENTARY INFORMATION}

\author{Jonah S. Peter}
\email{jonahpeter@g.harvard.edu}
\affiliation{Jet Propulsion Laboratory, California Institute of Technology, Pasadena, California 91109, USA}
\affiliation{Biophysics Program, Harvard University, Boston, Massachusetts 02115, USA}
\author{Tom A. Nordheim}
\affiliation{Jet Propulsion Laboratory, California Institute of Technology, Pasadena, California 91109, USA}
\author{Kevin P. Hand}
\affiliation{Jet Propulsion Laboratory, California Institute of Technology, Pasadena, California 91109, USA}

\maketitle

\renewcommand{\bibnumfmt}[1]{[S#1]}
\renewcommand{\citenumfont}[1]{S#1}

\section{Supplementary Results\label{sect:result}}

\subsection{Modeling landscape}

\begin{figure*}[htbp]
\centering 
\renewcommand\figurename{Supplementary Fig.}
\includegraphics[width=\textwidth]{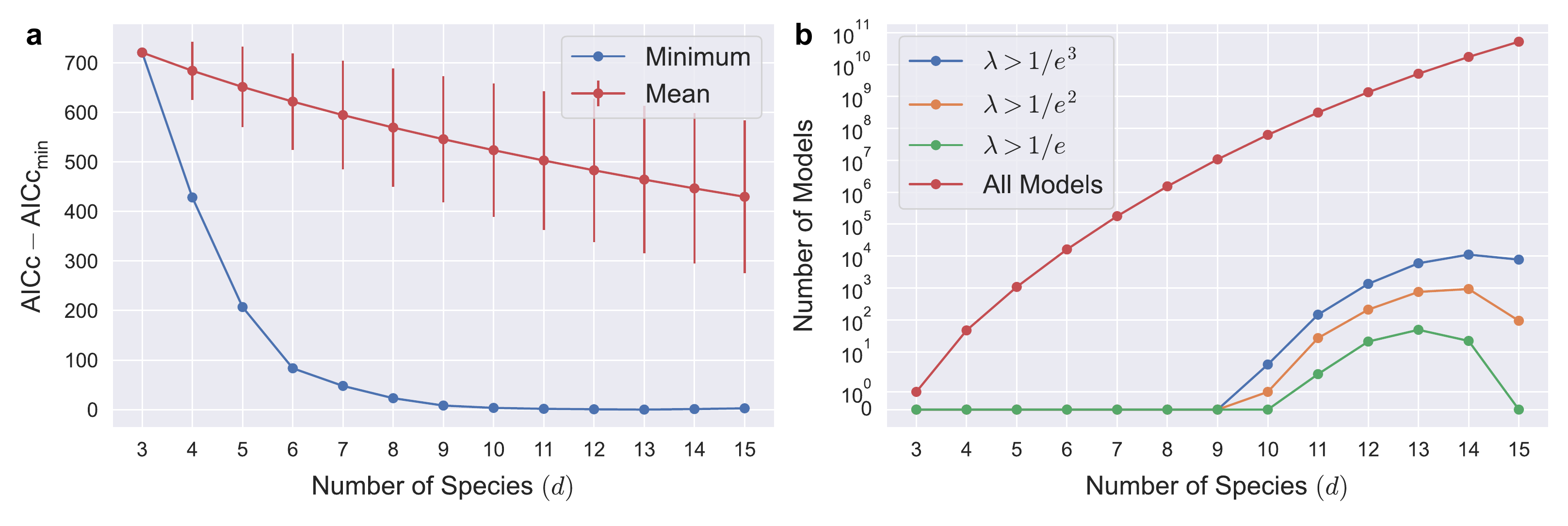}
\caption{\label{fig:landscape}Overview of the modeling landscape explored during the exhaustive model search. (a) The minimum (blue) and mean (red) AICc values across all models for each value of $d$ species. Vertical bars indicate $\pm 1$ standard deviation range. Note that the minimum AICc increases significantly from $d=13$ to 15 but appears flat on this axis scale (see Fig. 2a of the main text). (b) The number of models with $\lambda > 1/e^3$ (blue), $\lambda > 1/e^2$ (orange), and $\lambda > 1/e$ (green) compared to the total number of models (red) for each value of $d$.}
\end{figure*}

The exhaustive model search described in the main text consists of every possible species combination from $d=3$ to 15 and covers nearly 80 billion possible plume models. However, most of these models do not accurately reproduce the observed INMS data. Owing to the combinatorially large parameter space, only $\sim$1 in $10^9$ models exhibits strong predictive power. Supplementary Fig.~\ref{fig:landscape} demonstrates that the average model performs drastically worse than the optimal model for each value of $d$. This performance difference is significant across multiple standard deviations. In other words, although each species in the spectral library is a plausible plume constituent a priori, a random selection of compounds from the library would very likely not be capable of fitting the INMS data. Interestingly, the mean AICc decreases monotonically over the entire parameter range while the minimum AICc increases from $d=13$ to 15 (see also Fig. 2a of the main text). This trend demonstrates that the inclusion of additional species can appear to improve model performance indefinitely if only low likelihood models are considered. Such modeling behavior highlights the importance of exploring the entire parameter space in order to obtain accurate results (see also Supplementary Section~\ref{sect:method}).

\subsection{Alternative model selection criteria\label{sect:alt}}

\begin{table*}[htbp]
\centering
\begin{threeparttable}
\renewcommand\tablename{Supplementary Table}
\setcounter{table}{0}
\caption{\label{tab:alt_data}Modeling results for alternative model selection criteria. Mean values and uncertainties are presented as in the main text.}
\begin{tabular*}{\textwidth}{lllllll}
\toprule
& \multicolumn{2}{c}{$\lambda > 1/e$} & \multicolumn{2}{c}{$\lambda > 1/e^2$} & \multicolumn{2}{c}{Reduced ($\lambda > 1/e^3$)} \\
\cline{2-3} \cline{4-5} \cline{6-7} \\
Species & Probability & Mixing Ratio (\%) & Probability & Mixing Ratio (\%) & Probability & Mixing Ratio (\%)\\
\midrule
CH$_4$	                & 1    & $0.11 \pm 0.02$              &	1    & $0.11 \pm 0.02$              & 1    & $0.11 \pm 0.02$ \\
NH$_3$	                & 1    & $1.8 \pm 0.1$                & 1    & $1.8 \pm 0.1$                & 1    & $1.8 \pm 0.1$ \\
H$_2$O	                & 1    & $95.9 \pm 0.3$               & 1    & $95.9 \pm 0.3$               & 1    & $95.9 \pm 0.3$ \\
C$_2$H$_2$	            & 1    & $0.025 \pm 0.003$            & 0.95 & $0.024 \pm 0.004$            & 0.78 & $0.020 \pm 0.009$ \\
HCN	                    & 1    & $0.11 \pm 0.02$              & 1    & $0.11 \pm 0.02$              & 1    & $0.10 \pm 0.02$ \\
C$_2$H$_4$	            & 0    & $<6 \times 10^{-4}$\tnote{a} &	0.06 & $< 0.03$                     & 0.22 & $< 0.1$ \\
CO	                    & 1    & $0.73 \pm 0.06$              &	1    & $0.73 \pm 0.07$              & 1    & $0.72 \pm 0.07$\\
N$_2$	                & 0    & $<6 \times 10^{-4}$\tnote{a} & 0.01 & $< 0.002$                    & 0    & $<6 \times 10^{-4}$\tnote{a} \\
CH$_2$O	                & 0    & $< 0.001$\tnote{a}           & 0.03 & $< 0.002$                    & 0.07 & $< 0.005$ \\
NO	                    & 0.01 & $<7 \times 10^{-4}$\tnote{a} & 0.03 & $<0.001$                     & 0.12 & $<0.006$ \\
C$_2$H$_6$           	& 0.81 & $0.016 \pm 0.008$            & 0.74 & $0.014 \pm 0.009$            & 0.57 & $0.011 \pm 0.010$\\
CH$_5$N             	& 0.07 & $< 0.001$                    & 0.10 & $<0.002$                     & 0.22 & $< 0.005$ \\
CH$_3$OH            	& 0.49 & $< 0.005$                    & 0.48 & $<0.005$                     & 0.42 & $<0.005$ \\
H$_2$S              	& 0.38 & $< 0.003$                    & 0.39 & $< 0.003$                    & 0.33 & $< 0.002$\\
PH$_3$               	& 0.19 & $<0.001$                     & 0.22 & $<0.002$                     & 0.18 & $<0.001$\\
O$_2$	                & 0.59 & $<0.007$                     & 0.60 & $<0.008$                     & 0.49 & $<0.007$\\
$^{36}$Ar	            & 0    & $<4 \times 10^{-4}$\tnote{a} & 0.02 & $<4 \times 10^{-4}$\tnote{a} & 0    & $<4 \times 10^{-4}$\tnote{a}\\
C$_3$H$_4$           	& 0.05 & $<8 \times 10^{-4}$\tnote{a} &	0.10 & $<0.001$                     & 0.08 & $<0.001$ \\
$^{40}$Ar	            & 0.76 & $0.0012 \pm 0.0010$          &	0.66 & $0.0011 \pm 0.0010$          & 0.58 & $<0.004$ \\
CH$_3$CN	            & 0.07 & $<0.001$                     & 0.11 & $< 0.002$                    & 0.07 & $< 0.002$\\
C$_2$H$_2$O	            & 0    & $<7 \times 10^{-4}$\tnote{a} & 0.01 & $<7 \times 10^{-4}$\tnote{a} & 0    & $<7 \times 10^{-4}$\tnote{a} \\
C$_3$H$_6$	            & 1    & $0.005 \pm 0.002$          & 0.92 & $0.004 \pm 0.002$              & 0.94 & $0.005 \pm 0.002$\\
CH$_2$N$_2$	            & 0    & $<4 \times 10^{-4}$\tnote{a} & 0.04 & $<4 \times 10^{-4}$\tnote{a} & 0.03 & $<4 \times 10^{-4}$\tnote{a}\\
C$_2$H$_4$O	            & 0    & $<7 \times 10^{-4}$\tnote{a} &	0.05 & $< 0.001$                    & 0.12 & $<0.003$\\
C$_3$H$_8$	            & 0.19 & $<0.005$                     & 0.22 & $< 0.005$                    & 0.22 & $<0.006$\\
CO$_2$	                & 1    & $0.45 \pm 0.03$              &	1    & $0.45 \pm 0.03$              & 1    & $0.45 \pm 0.03$\\
C$_2$H$_7$N         	& 0    & $<0.002$\tnote{a}            & 0.01 & $<0.002$\tnote{a}            & 0    & $<0.002$\tnote{a} \\
CH$_3$NO	            & 0    & $<5 \times 10^{-4}$\tnote{a} &	0.01 & $<5 \times 10^{-4}$\tnote{a} & 0    & $<5 \times 10^{-4}$\tnote{a} \\
C$_2$H$_6$O          	& 0    & $<0.002$\tnote{a}            & 0.01 & $<0.002$\tnote{a}            & 0    & $<0.002$\tnote{a} \\
CH$_2$O$_2$          	& 0    & $<7 \times 10^{-4}$\tnote{a} & 0.01 & $<7 \times 10^{-4}$\tnote{a} & 0    & $<7 \times 10^{-4}$\tnote{a} \\
C$_4$H$_8$	            & 0    & $<7 \times 10^{-4}$\tnote{a} & 0.01 & $<7 \times 10^{-4}$\tnote{a} & 0    & $<7 \times 10^{-4}$\tnote{a} \\
C$_2$H$_6$N$_2$      	& 0.38 & $<0.002$                     & 0.29 & $< 0.002$                    & 0.12 & $<9 \times 10^{-4}$\\
C$_3$H$_6$O	            & 0.23 & $<0.001$                     & 0.17 & $<0.001$                     & 0.10 & $<7 \times 10^{-4}$\\
C$_4$H$_{10}$	        & 0.04 & $<0.008$\tnote{a}            & 0.08 & $<0.008$\tnote{a}            & 0.07 & $<0.008$\tnote{a}\\
C$_2$H$_4$O$_2$      	& 0    & $<6 \times 10^{-4}$\tnote{a} & 0.01 & $<6 \times 10^{-4}$\tnote{a} & 0 & $<6 \times 10^{-4}$\tnote{a} \\
C$_3$H$_8$O	            & 0.06 & $<0.003$\tnote{a}            & 0.09 & $<0.004$                     & 0.02 & $<0.003$\tnote{a} \\
C$_2$H$_7$NO	        & 0.16 & $<0.002$                     & 0.21 & $< 0.002$                    & 0.19 & $< 0.002$\\
C$_2$H$_6$O$_2$	        & 0.23 & $<0.01$                      &	0.21 & $< 0.01$                     & 0.04 & $<0.005$\tnote{a}\\
C$_5$H$_{10}$       	& 0    & $<6 \times 10^{-4}$\tnote{a} & 0.01 & $<6 \times 10^{-4}$\tnote{a} & 0    & $<6 \times 10^{-4}$\tnote{a}\\
C$_4$H$_9$N	            & 0.02 & $<2 \times 10^{-4}$\tnote{a} & 0.04 & $<2 \times 10^{-4}$\tnote{a} & 0.05 & $<2 \times 10^{-4}$\tnote{a}\\
C$_5$H$_{12}$	        & 0.02 & $<0.002$\tnote{a}            & 0.05 & $<0.002$\tnote{a}            & 0.05 & $<0.002$\tnote{a}\\
C$_4$H$_{10}$O	        & 0    & $<4 \times 10^{-4}$\tnote{a} & 0.02 & $<4 \times 10^{-4}$\tnote{a} & 0.01 & $<4 \times 10^{-4}$\tnote{a}\\
C$_2$H$_5$NO$_2$	    & 0.02 & $<0.007$                     & 0.04 & $< 0.01$                     & 0.08 & $<0.02$\\
C$_3$H$_5$Cl	        & 0.08 & $< 4 \times 10^{-4}$         & 0.18 & $< 0.001$                    & 0.02 & $<2 \times 10^{-4}$\tnote{a}\\
C$_5$H$_9$N	            & 0.02 & $<3 \times 10^{-4}$\tnote{a} & 0.10 & $<3 \times 10^{-4}$\tnote{a} & 0.03 & $<3 \times 10^{-4}$\tnote{a}\\
C$_4$H$_6$O$_2$      	& 0.05 & $<6 \times 10^{-4}$\tnote{a} & 0.07 & $<6 \times 10^{-4}$\tnote{a} & 0.07 & $<6 \times 10^{-4}$\tnote{a}\\
C$_3$H$_7$NO$_2$	    & 0    & $<1 \times 10^{-4}$\tnote{a} & 0.02 & $<3 \times 10^{-4}$          & 0    & $<1 \times 10^{-4}$\tnote{a} \\
C$_8$H$_{18}$	        & 0.02 & $<1 \times 10^{-4}$\tnote{a} & 0.03 & $<1 \times 10^{-4}$\tnote{a} & 0.04 & $<1 \times 10^{-4}$\tnote{a}\\
C$_6$H$_{12}$N$_4$	    & 0    & $<5 \times 10^{-5}$\tnote{a} & 0.03 & $<5 \times 10^{-5}$\tnote{a} & 0    & $<5 \times 10^{-5}$\tnote{a}\\
C$_6$H$_{14}$N$_2$O$_2$	& 0    & $<1 \times 10^{-4}$\tnote{a} & 0.01 & $<1 \times 10^{-4}$\tnote{a} & 0 & $<1 \times 10^{-4}$\tnote{a} \\
\bottomrule
\end{tabular*}
\begin{tablenotes}
\item[a]{Denotes number density relative to H$_2$O.}
\end{tablenotes}
\end{threeparttable}
\end{table*}

As discussed in the main text, it is advisable to limit the total number of candidate models in order to make reliable statistical inferences. This consideration motivates the use of a restricted spectral library consisting of only geochemically plausible (and INMS detectable) compounds, as well as the elimination of extraneous low likelihood models. The criteria used to define which models are ``high likelihood" requires a tradeoff between neglecting alternative explanations (i.e., bias; underfitting) and risking parameter estimation based on poor performing models (i.e., variance; overfitting). The inferences of the main text are based on the exhaustive list of models satisfying $\lambda > 1/e^3$. This threshold was chosen to reduce bias and because Monte Carlo simulations suggest that it corresponds to a $\sim$95--99\% model confidence interval~\cite{burnham_model_2004, richards_dealing_2007, richards_model_2011, richards_testing_2005}. To assess the influence of this value on our results, we repeated the analysis of the main text for the increasingly restrictive thresholds $\lambda > 1/e^2$ and $\lambda > 1/e$. Correspondingly, results based on these criteria are made using an increasingly small set of only the highest likelihood models (see also Fig. 2a of the main text). In both cases, our results remain virtually unchanged (Supplementary Table~\ref{tab:alt_data}). The use of a more restrictive threshold leads to slightly more precise mixing ratios for some minor compounds, as expected with a bias-variance tradeoff. Notably, even the lowest likelihood species listed in Table 1 of the main text are also contained within the more restricted sets of models. This result suggests that adopting the lower (more unbiased) threshold of $\lambda > 1/e^3$ does not lead to overfitting as a result of the increased variance.

The decision to use the exhaustive model list was made to further reduce model selection bias as much as possible. We believe this is appropriate given the limited data and large number of plausible model components~\cite{symonds_brief_2011}. Nevertheless, alternative procedures could be implemented by factoring in prior knowledge about the INMS data set. To investigate the stability of our results under different model selection criteria, we repeated the analysis once more by applying the following additional selection criteria to the initial set of $\lambda > 1/e^3$ models. First, we excluded all models containing more than one species with mass greater than 50 u, above which INMS did not detect any signal above background noise~\cite{waite_cassini_2017, postberg_plume_2018}. Doing so conserves statistical power in an effort to focus on lower mass compounds for which there is prior circumstantial evidence~\cite{waite_cassini_2006, waite_jr_liquid_2009}. Next, in order to reduce redundancy, we eliminated models with any pairwise regressor correlations greater than $\rho = 0.9$~\cite{james_introduction_2013, burnham_model_2004, symonds_brief_2011} (see also Supplementary Section~\ref{sect:pairwise} below). Models with $d>13$ were also excluded for the same purpose (see Fig. 2b of the main text). Finally, if there existed a simpler model with a lower AICc that was fully nested within a more complex model, the more complex model was discarded. In this scenario, the more complex model performs well only because it contains all the parameters of the simpler model, and some studies suggest that this rule improves model selection accuracy by reducing the chance of false positive results~\cite{burnham_model_2004, symonds_brief_2011, richards_model_2011, richards_dealing_2007}. The species probabilities and mixing ratios determined on the reduced set of models are shown in the rightmost column of Supplementary Table~\ref{tab:alt_data}. Once again, there is very good agreement between these results and those of the main text, further indicating that the reduction in model bias imposed by the exhaustive search is not associated with overfitting.

 Overall, the strong similarity of the results across multiple different selection procedures demonstrates stability in the modeling process. Nevertheless, we recommend that quantitative inferences be made using the values presented in the main text. Those results represent the most unbiased (and conservative) determination of the plume's composition. By contrast, the increased precision of the results listed in Supplementary Table~\ref{tab:alt_data} may not be warranted and should be interpreted only as supporting evidence.
 
\subsection{Analysis of pairwise collinearity\label{sect:pairwise}}

Compositional ambiguities within INMS spectra arise due to the large number of candidate species combined with the relatively low mass resolution of the instrument. In other words, models of the plume's composition contain many possible model components with comparatively few data points available to constrain them. This difficulty is accentuated when there exist combinations of multiple species that can reproduce the signal produced by another. If this collinearity is exact, discriminating between such singular species in the composite INMS spectrum is impossible. In practice, even approximately collinear mass spectra can present a significant obstacle when interpreting data with a finite mass resolution. These issues have been encountered in previous studies~\cite{postberg_plume_2018, waite_cassini_2006, waite_jr_liquid_2009} and have greatly hindered efforts to identify trace compounds in the plume.

Here, we quantify the extent of pairwise collinearity amongst library spectra by computing the correlation matrix for each species’ cracking pattern. We use $\rho$ to denote correlation coefficients of cracking patterns(i.e., the regressors themselves), in contrast to $r$ (defined in the main text), which signifies correlations between regression coefficients. Whereas $r$ is a model dependent quantity (see Fig. 2b of the main text), $\rho$ is a static property of the spectral library. 

\begin{figure*}[tpb]
\centering 
\renewcommand\figurename{Supplementary Fig.}
\includegraphics[width=\textwidth]{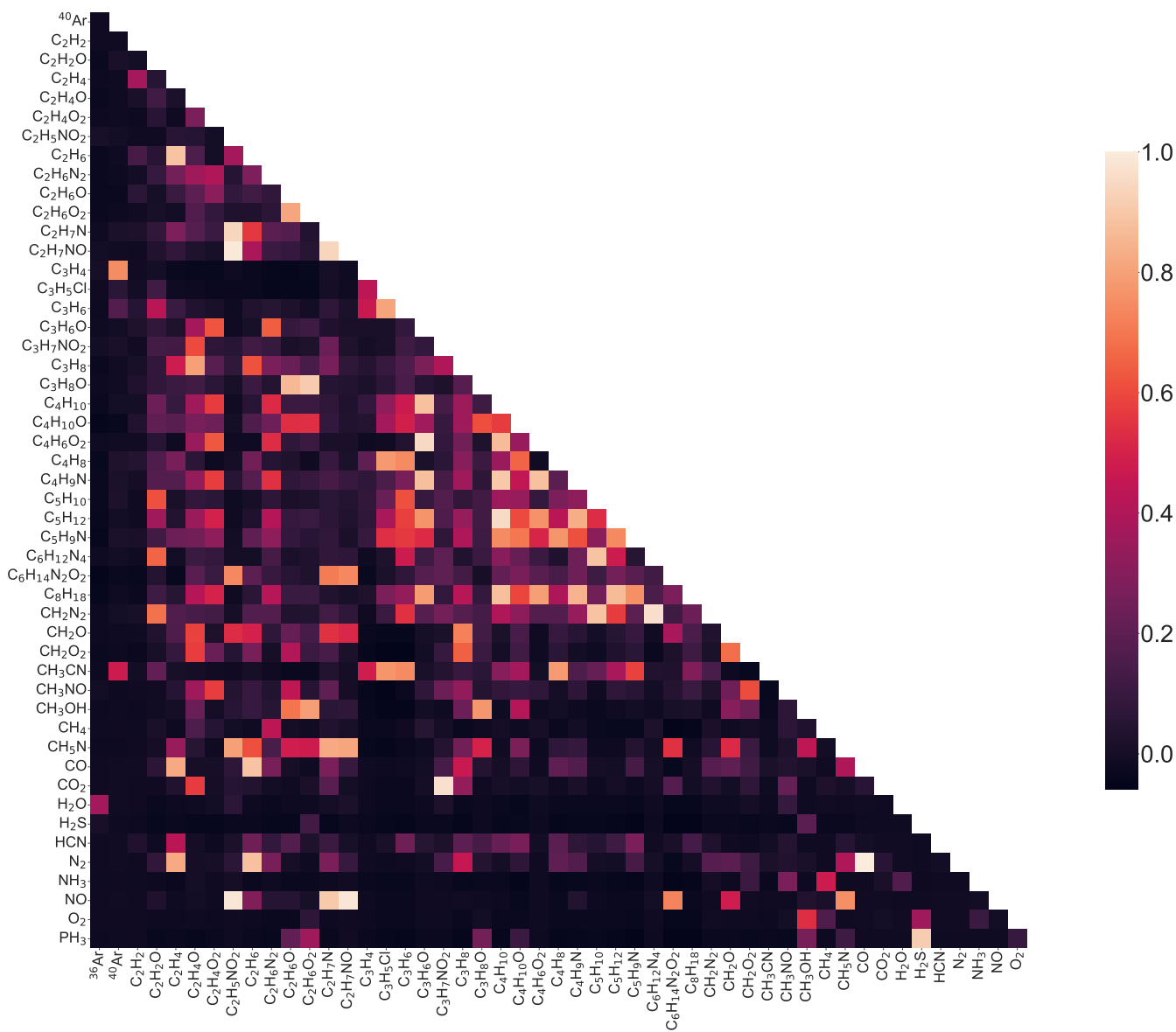}
\caption{\label{sfig1}Correlation matrix for the spectral library. The correlation coefficient ($-1 \leq \rho \leq 1$) measures the extent of collinearity between two cracking patterns. Large positive values (brighter colors) indicate high levels of collinearity, whereas smaller values signify less collinearity. Large negative values (darker colors) are essentially absent, indicating that most collinear species are positively correlated. }
\end{figure*}

Supplementary Fig.~\ref{sfig1} demonstrates that large positive correlations are common and manifest between molecules with similar cracking patterns. A correlation coefficient between two species of $\rho=1$ would imply identical mass spectra. Pairs of species with correlation coefficients close to 1 are approximately collinear and may still be indistinguishable in practice. By contrast, pairs of species with correlation coefficients close to 0 lack a consistent pattern of strong overlapping features in their cracking patterns. Values near 0.5 represent the intermediate case where two species share certain features but also possess additional large mass peaks that are not shared. NH$_3$ and CH$_4$, for example, have a correlation coefficient of 0.47, owing to their shared peak at mass 16 and unshared peaks at masses 17 (NH$_3$) and 15 (CH$_4$). Of course, large negative values are also possible, though they are effectively absent from the spectral library due to the inherent structural similarities between organic compounds. Large negative values would signify that one species has significant mass peaks predominantly at mass channels where another species does not. 

\begin{figure}
\centering 
\renewcommand\figurename{Supplementary Fig.}
\includegraphics[width=\textwidth]{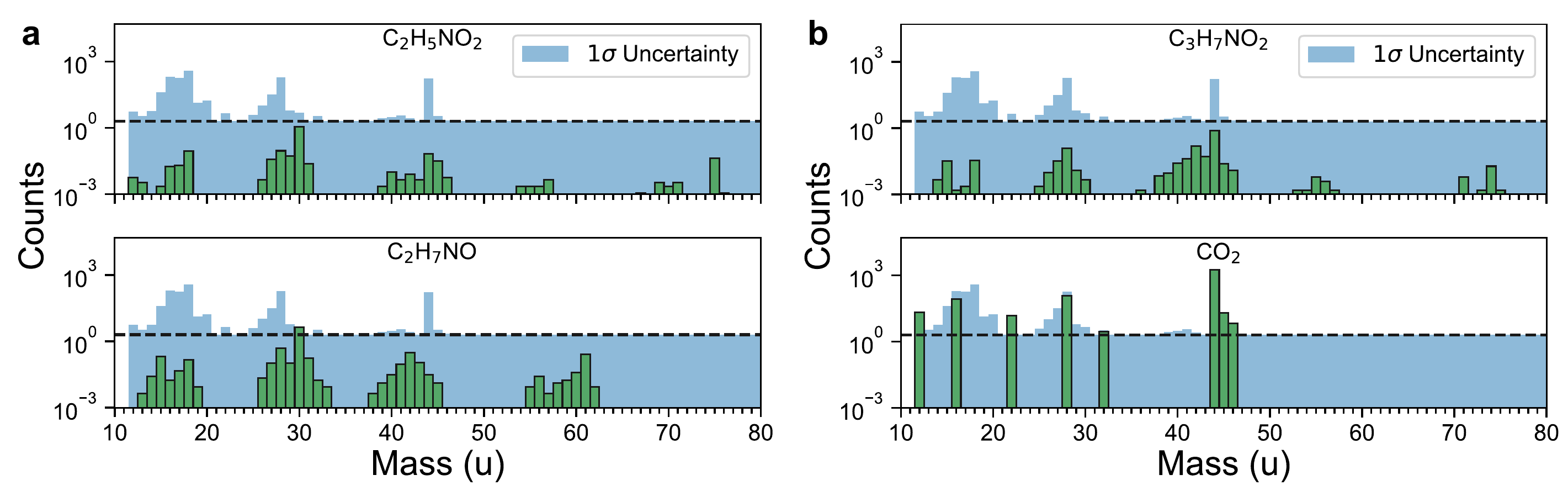}
\caption{\label{fig:amino} (a) Mass spectra for glycine and C$_2$H$_7$NO. The x-axis shows the mass of each fragmentation product, while the y-axis quantifies the proportional abundance of each fragment based on the mean mixing ratios determined in the main text (note the log-scale). All spectral features (green bars) are masked by the associated count uncertainty at each mass channel (shaded blue region). Both species have a single dominant peak at 30 u, leading to high pairwise correlations. Only the signal from C$_2$H$_7$NO extends above the minimum count uncertainty (dashed black line). (b) Mass spectra for alanine and CO$_2$. Both species share a single dominant peak at mass 44.}
\end{figure}

For compounds with cracking patterns dominated by only a few major peaks, sharing as few as one of these peaks can lead to high correlation coefficients. A notable example is C$_2$H$_7$NO and glycine (C$_2$H$_5$NO$_2$), which exhibit a correlation coefficient of $\rho = 0.99$. Based on the results of the main text, it is tempting to view the presence of glycine in 2 of the 94 highest likelihood models ($\lambda > 1/e$) as the first tentative evidence for amino acids at Enceladus. However, at such low concentrations, glycine is virtually indistinguishable from C$_2$H$_7$NO. Although the cracking pattern for glycine contains counts at many different mass channels, the peak at 30 u is by far the dominant feature. As a result, this mass channel acts as a high leverage point and drags up correlations between glycine and other species with large peaks at 30 u. This effect is particularly pronounced when there are no other major peaks present, as is the case for C$_2$H$_7$NO. Supplementary Fig.~\ref{fig:amino}a shows the mean contributions of glycine and C$_2$H$_7$NO to the INMS spectrum calculated based on the multi-model averaging procedure presented in the main text. All peaks are well within the associated $1 \sigma$ uncertainty for each mass channel. The base peak is the only C$_2$H$_7$NO feature that extends above the minimum count uncertainty and mimics the signal for glycine at similar concentrations. This ambiguity precludes the detection of trace amounts of glycine in the plume and might have important implications for the detection of amino acids on future spacecraft missions as well. The high pairwise correlation between glycine and C$_2$H$_7$NO suggests that glycine likely cannot be independently detected at Enceladus using a 1 u resolution mass spectrometer (such as INMS). Instead, an independent measurement of the C$_2$H$_7$NO mixing ratio with precision at least as great as the instrument used to detect glycine would be necessary. A similar argument applies to alanine (C$_3$H$_7$NO$_2$) and CO$_2$ ($\rho = 0.97$) which share a single dominant peak at mass 44 (Supplementary Fig.~\ref{fig:amino}b).

Ambiguities of this nature also underlie the increased mixing ratio uncertainties deduced for the alcohols and mass 43 fragments discussed in the main text. Large positive correlations amongst the alcohols (as high as $\rho = 0.91$) manifest via high count rates at 31 u, corresponding to the hydroxymethyl group, CH$_2$OH$^+$. Species with prominent 43 u fragments exhibit correlations as high as $\rho = 0.95$ and could be attributable to a number of different structures (see Supplementary Fig.~\ref{similar} and Supplementary Section~\ref{sect:comp_res}). Consequently, inter-model uncertainty stemming from these correlations reduces the precision of the final averaged mixing ratio.

\begin{figure}[t]
\centering 
\renewcommand\figurename{Supplementary Fig.}
\includegraphics[width=0.65\textwidth]{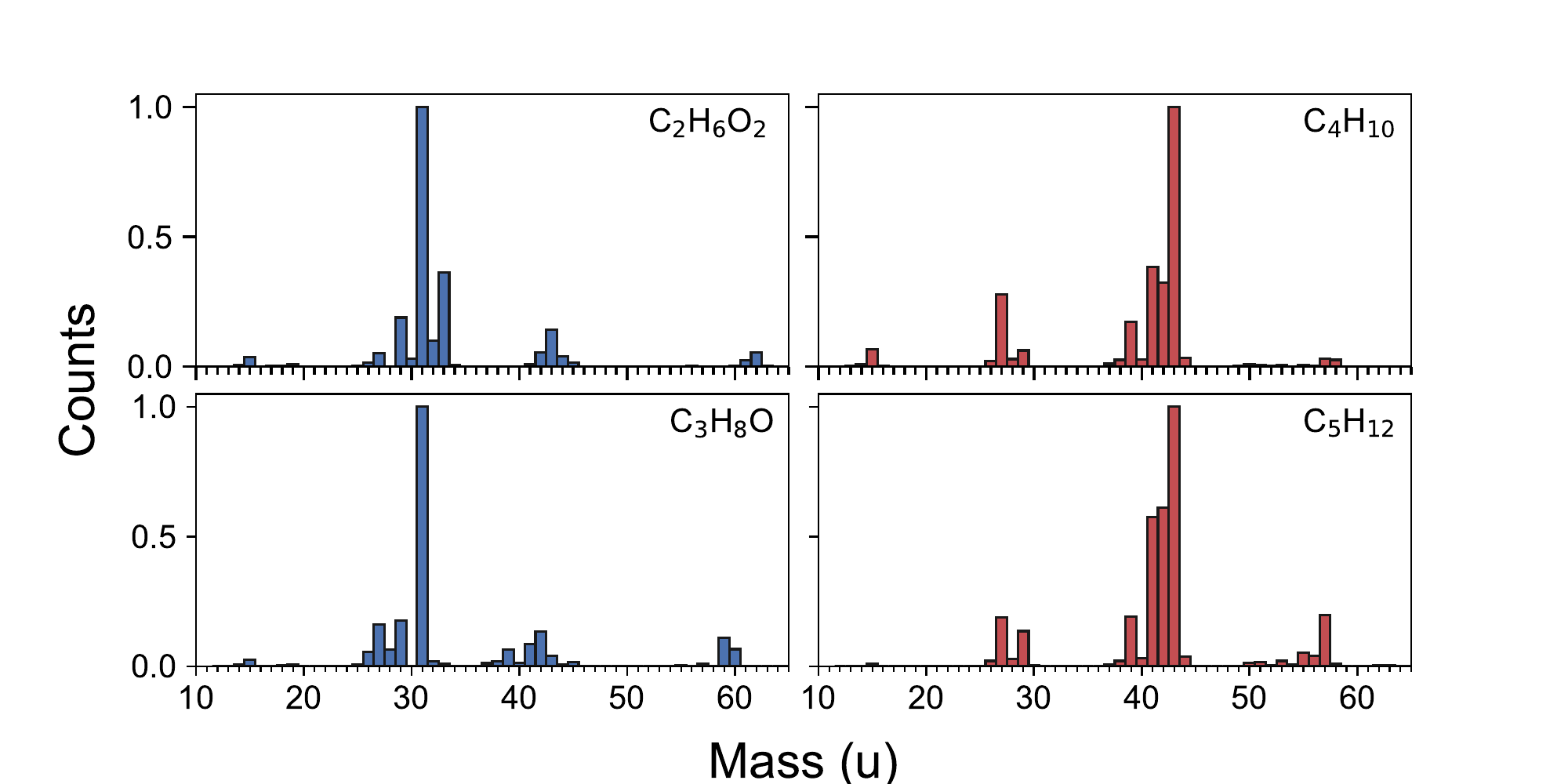}
\caption{\label{similar}Mass spectra for representative alcohols and species with 43 u fragments. Blue bars (left) show representative alcohols, C$_2$H$_6$O$_2$ and C$_3$H$_8$O, with $\rho = 0.91$. Red bars (right) show representative species with strong 43 u signatures, C$_4$H$_{10}$ and C$_5$H$_{12}$, with $\rho = 0.95$.}
\end{figure}

\subsection{Ramifications for hypothesis testing}

Extensive amounts of collinearity between model components can have profound consequences on statistical inference. Although one might expect traditional hypothesis testing to identify important model parameters, high dimensionality and collinearity of the spectral library limit statistical power and prevent individual regression coefficients from achieving statistical significance. Supplementary Table~\ref{tab:hypothesis} demonstrates this phenomenon for the low velocity INMS data set. An $F$-test for overall significance indicates that the spectral library, in aggregate, does indeed better explain the INMS data than does the ``null model,” which consists of fitting only the regression coefficient for H$_2$O ($F = 12$; $p = 1.8 \times 10^{-14}$). However, one-tailed $t$-tests indicate that only the three strongest spectral features, H$_2$O, NH$_3$, and CO$_2$ are individually statistically significant ($p = 2.6 \times 10^{-79}$, $6.0 \times 10^{-16}$, and $9.5 \times 10^{-7}$, respectively) in the presence of the entire spectral library. These species account for the majority of counts at masses 16, 17, 18, and 44 and produce signals that are large enough to stand out amongst the majority of collinear species that comprise the rest of the library. Though we can be confident that the aggregate set of candidate library species is capable of explaining the INMS data, the statistical significance of any one species is difficult to show using such frequentist statistics under these conditions. For this reason, information theoretic approaches (such as those described in the main text) or related Bayesian analyses are more suitable for INMS spectra deconvolution.

\begin{table}[h]
\centering 
\renewcommand\tablename{Supplementary Table}
\begin{threeparttable}
\caption{\label{tab:hypothesis}Results of hypothesis testing on the INMS data set. High dimensionality and collinearity prevent all but the most prominent spectral features from achieving statistical significance.}
\begin{tabularx}{0.5\textwidth}{lll}
\toprule
Species & $t$-Statistic  & $p$-Value \\
\midrule
H$_2$O & $380$ & $2.6 \times 10^{-79}$\\
NH$_3$ & $12$  & $6.0 \times 10^{-16}$\\
CO$_2$ & $5.5$ & $9.5 \times 10^{-7}$\\
All others & $< 1.4$ & $> 0.08$\\
\bottomrule
\end{tabularx}
\end{threeparttable}
\end{table}

\section{Supplementary Discussion\label{sect:discuss}}

\subsection{Comparison to other results\label{sect:comp_res}}

It is important to consider the entire body of statistical evidence when drawing conclusions about which species are detected in the plume. The model-averaged mixing ratios, the minimum AICc model, the individual species probabilities, and the relative model likelihoods can all be used to assess the confidence of each detection. The more heuristics that point towards the detection of a given species, the more confident one may be that said species is truly present in the plume.  

The minimum AICc model consists of 13 species. One interpretation of this result is to classify all 13 species as conclusive detections. However, more parsimonious models exist that explain the INMS data nearly as well. A more conservative approach would be to interpret only the species that comprise the best-fitting, least complex model with $\lambda > 1/e^3$ as essential to the fit. We favor an even more nuanced approach and suggest using a tiered hierarchy of confidence based on the holistic analysis of the main text. Below, we contextualize these results within the collection of previous studies on the composition of Enceladus and other icy bodies.

In the main text, we present an NH$_3$ mixing ratio that is larger than what has been reported based on previous analyses of the slow flyby INMS spectra~\cite{waite_cassini_2017, postberg_plume_2018}, as well as those obtained during the E2~\cite{waite_cassini_2006} and E3~\cite{waite_jr_liquid_2009} flybys. In their supplementary material, Waite et al.~\cite{waite_cassini_2017} argue, based on the slow flyby data, that the residual left at mass 16 by fitting H$_2$O, CH$_4$, and CO$_2$ alone unambiguously indicates the presence of NH$_3$. For the spectrum analyzed in the main text of this work, we find that this residual is equal to $\sim$1330 counts. The NH$_3$~$/$~H$_2$O number density calculated based on this residual alone is $\sim$0.01, in agreement with ref.~\cite{waite_cassini_2017}. Of course, this conclusion is dependent on the estimation of count uncertainties in INMS data, for which multiple different procedures have been proposed~\cite{waite_jr_liquid_2009, waite_cassini_2017, magee_inms-derived_2009, cui_analysis_2009}. The precise uncertainties adopted in ref.~\cite{waite_cassini_2017} may not be identical to those used here, and this difference could be a source of diverging results. Telescopic data have suggested the presence of NH$_3$ or NH$_3$-hydrate~\cite{zastrow_uv_2012, grundy_near-infrared_1999, emery_near-infrared_2005, verbiscer_near-infrared_2006} at Enceladus, though these observations are not definitive~\cite{cruikshank_spectroscopic_2005} and other instruments including the \textit{Cassini} Visible and Infrared Mapping Spectrometer (VIMS), Ultraviolet Imaging Spectrograph (UVIS), and Cosmic Dust Analyzer (CDA) have failed to find conclusive evidence for its presence in the plume~\cite{brown_composition_2006, hansen_composition_2020, postberg_plume_2018}. Although we agree with the authors of ref.~\cite{waite_cassini_2017} that NH$_3$ is required to explain the INMS data, additional studies that further examine the effects of uncertainty estimation would be helpful in validating the precise mixing ratio.

As discussed in the main text, suggestive evidence for nitrogen at Enceladus in the form of HCN has been previously reported based on other INMS data sets. HCN has also been suggested to help explain yet unresolved signatures in Cassini CDA spectra of ice grains in Saturn’s E-ring~\cite{hillier_composition_2007}, though could not be definitively identified by CDA as a cation species due to its high ionization potential~\cite{postberg_e-ring_2008}. As such, the HCN mixing ratio determined in this work is the first definitive detection of nitrile chemistry at Enceladus.

Our analysis shows that C$_2$H$_2$ and C$_2$H$_6$ are also present in the plume with concentrations exceeding 100 ppm. As for HCN, both species have been detected in comets~\cite{mumma_chemical_2011} and circumstantial evidence for their existence at Enceladus has been previously suggested based on other Cassini flybys. C$_2$H$_2$ was suspected by Waite et al.~\cite{waite_jr_liquid_2009} during the high-velocity E3 and E5 flybys, though correlations between abundance and spacecraft velocity suggest that impact fragmentation may have been responsible. Only upper limits for C$_2$H$_6$ were reported based on the E2, E3, and E5 flybys~\cite{waite_jr_liquid_2009}. The presence of C$_2$H$_6$ would also help explain features at 28--30 u seen in CDA ice grain spectra~\cite{postberg_plume_2018}, which are consistent with the INMS data.

Concerning the higher mass organics, we find that trace amounts ($\sim$40 ppm) of C$_3$H$_6$ (and possibly C$_3$H$_8$) are present in the plume as well. C$_3$H$_6$ was produced as a fragmentation product during the E3 and E5 flybys of Enceladus~\cite{waite_jr_liquid_2009}, and both species have been observed at Titan~\cite{magee_inms-derived_2009}. Although not initially identified in the E2 flyby data~\cite{waite_cassini_2006}, reanalysis suggests that C$_3$H$_6$ may have been present~\cite{waite_jr_liquid_2009}; however it was not identified by Waite et al. as an intrinsic plume constituent in subsequent publications~\cite{waite_cassini_2017, postberg_plume_2018}. Our analysis shows that C$_3$H$_6$ is by far the most likely explanation for the count signal in the 37--42 u region of the slow flyby INMS spectrum (Fig. 4d of the main text) and provides the first conclusive identification of native C$_3$ organics in the plume.  

Native alcohols have not been previously identified in the plume, but CH$_3$OH has been suggested as a possible fragmentation product based on high velocity INMS and ice grain spectra~\cite{postberg_macromolecular_2018}. CH$_3$OH has also been observed via ground-based methods in the vicinity of Enceladus~\cite{drabek-maunder_ground-based_2019} and has been proposed as an explanation for the 3.53 $\mu$m absorption feature seen in VIMS spectra of the surface~\cite{hodyss_methanol_2009}. CH$_3$OH could be produced through chemical processing of CH$_4$ in the nearby gas cloud  or by the partial combustion of endogenous CH$_4$ within the ocean. Both CH$_3$OH and C$_2$H$_6$O$_2$ have been detected in relatively high abundance in several comets~\cite{biver_complex_2014, crovisier_ethylene_2004}.

As discussed in the main text, the authors of ref.~\cite{waite_cassini_2017} identified a possible instrument artifact at mass 32 that complicates the determination of the native plume O$_2$ abundance. In accounting for this artifact, they estimated a corrected mass 32 signal equal to $(100/45) \times 0.004\% = 0.0089\%$ of the counts measured at mass 18. The value of mass 32 counts used in this work is well within this limit, and the concomitant detections of multiple partially oxidized organics are consistent with a native source of O$_2$. However, the precise source of O$_2$ at Enceladus presents a few challenges. Unlike Europa, where charged particle bombardment of the surface is known to drive radiolysis of water and other elements to O$_2$, H$_2$O$_2$, SO$_4^{2-}$, and other oxidants~\cite{hand_physics_2007, hand_keck_2013, spencer_condensed_2002, trumbo_h_2019}, the radiation flux near Enceladus is considerably lower~\cite{paranicas_energetic_2012}, and evidence for oxidants on the surface is lacking, despite proposals of such radiolytic chemistry~\cite{ray_oxidation_2021, newman_hydrogen_2007, loeffler_is_2009}. Even with moderate levels of surface radiolysis, a key problem would be the efficiency of oxidant production at the high temperatures observed in the South Polar Terrain~\cite{hand_h2o2_2011, spencer_cassini_2006}. Here we do not propose a solution for this issue but rather note that our results are consistent with the presence of O$_2$, be it from surface radiolysis and subsequent delivery to the ocean, or via other production mechanisms. Notably, Waite et al.~\cite{waite_cassini_2017} found that radiolysis of H$_2$O due to radioactive isotopes in Enceladus’ core could also produce O$_2$ at the abundance reported here.

Evidence for $^{40}$Ar stems from the large $\sim$2.2$\sigma$ residual at 40 u, which is the only mass channel with strong enough signal to significantly influence the calculation of its mixing ratio. This signal does have strong overlap with the cracking pattern of C$_3$H$_4$ ($\rho = 0.75$), but the abundance of this species is limited by the larger contribution of C$_3$H$_6$ at neighboring mass channels. Though not identified in a previous analysis of the slow flybys~\cite{waite_cassini_2017}, $^{40}$Ar was detected during the E3 and E5 flybys and may indicate significant water-rock interactions and the leaching of salts within Enceladus~\cite{glein_geochemistry_2018, waite_jr_liquid_2009}. CH$_3$CN also has a reasonably high correlation with $^{40}$Ar ($\rho = 0.47$) and could potentially contribute to the observed signal at mass 40. Although there is no definitive evidence for CH$_3$CN at Enceladus, its presence would not be unexpected based on the evidence for HCN and hydrocarbons such as C$_3$H$_6$ (see Discussion in the main text).

Our results also provide strong evidence for an ambiguous species with a strong 43 u signal. Ambiguity at this mass channel was previously reported based on the E5 flyby of Enceladus~\cite{waite_jr_liquid_2009}. One explanation of this signal is C$_2$H$_6$N$_2$. Although this species has an additional large peak at 15 u, this feature is masked by the much larger contribution from CH$_4$ in the INMS data. The correlated ice grain features at 15 and 43 u seen in ``Type II” CDA spectra~\cite{postberg_e-ring_2008} might be explained by fragmentation of C$_2$H$_6$N$_2$ into CH$_3^+$ and CH$_3$N$_2^+$. Alternative explanations including acetyl group-bearing species such as C$_3$H$_6$O or C$_4$H$_6$O$_2$ are also possible.

Sulfur compounds have not been definitively identified at Enceladus, though a tentative past detection of H$_2$S based on the E5 flyby~\cite{waite_jr_liquid_2009, postberg_plume_2018} supports our finding that it may be present. H$_2$S would be expected if there is active serpentinization taking place on the ocean floor. PH$_3$ has not been previously reported in the plume, but may have been observed in the coma of comet 67P/Churyumov-Gerasimenko~\cite{altwegg_prebiotic_2016}. A recent analysis of salt-rich ice grains collected by CDA in Saturn's E-ring detected phosphorus in the form of Na$_3$PO$_4$ and Na$_2$HPO$_4$, which may imply an abundant source of phosphorus within Enceladus' ocean~\cite{postberg_detection_2023}.

For the remaining species listed in Table 1 and Extended Data Table 2 of the main text, prior evidence for their existence in the plume is relatively poor. Neither CH$_5$N nor C$_3$H$_5$Cl have been identified at Enceladus, but Cl has been detected by CDA as NaCl and KCl salts residing in plume ice grains~\cite{postberg_sodium_2009, postberg_salt-water_2011}. Lastly, although there is no strong evidence for amino acids at Enceladus (see Supplementary Section~\ref{sect:pairwise}), we note that glycine, alanine, and other amino acids are abundant in carbonaceous chondrites~\cite{glavin_search_2020}. 

\subsection{Comparison to other methodologies\label{sect:method}}

In order to adequately account for the high dimensionality and (approximate) collinearity of the INMS plume dataset, it is necessary to perform a type of variable selection that constrains the parameter space of possible model fits. Such variable selection techniques trade off a small increase in model bias for a significant reduction in model variability~\cite{guyon_introduction_2003, james_introduction_2013}. The resulting models are far less likely to overfit noisy features in the training data and tend to be significantly more accurate in predicting future observations~\cite{james_introduction_2013}. Moreover, variable selection reduces the impact of collinearity by identifying which model components are better at explaining the observed data and discarding those that are superfluous. Such a process allows for the evaluation of individual model components without the confounding presence of their collinear counterparts. Subsequent multi-model inference based on model averaging then ensures that all candidate models are considered on equal footing.

In the main text, we outlined two variable selection procedures: an exhaustive all subset selection for models with fewer than 16 species and a forward stepwise selection algorithm for more complex models. Additional model selection criteria are considered in Supplementary Section~\ref{sect:alt} above. Other common algorithms such as ridge regression~\cite{hoerl_ridge_1970} and the Least Absolute Shrinkage and Selection Operator (LASSO) are frequently applied in a wide variety of machine learning and model validation contexts~\cite{guyon_introduction_2003, hocking_biometrics_1976, james_introduction_2013, santosa_linear_1986, tibshirani_regression_1996}. These methods seek to reduce the influence of extraneous parameters via L2 or L1 regularization, respectively. Multi-model averaging is similar to L1 regularization in that it allows for explicit dimensionality reduction, whereas L2 regularization does not. This property leads to high model interpretability, which is of utmost importance when performing compositional analyses. Other heuristics besides the AICc can be used to select models for averaging, but these alternative statistics are not based on minimizing information loss and are therefore not well-suited for model selection when the structure of the unknown, true distribution is poorly constrained. The property of the AICc as a minimum variance unbiased estimator of model performance supports its use in this context~\cite{davies_estimation_2006}. Furthermore, model inference using the AICc has been shown to asymptotically approximate results based on cross-validation (another broadly accepted model validation technique) while requiring much fewer computational resources~\cite{stone_asymptotic_1977, stoica_model-structure_1986, burnham_model_2004}.

The first few studies of INMS data collected at Enceladus produced landmark results, including the characterization of major plume constituents and the detection of molecular H$_2$ as a potential indicator of hydrothermal activity~\cite{waite_cassini_2006, waite_jr_liquid_2009, waite_cassini_2017}. These papers (and related works published throughout the duration of the \textit{Cassini} mission) also documented a detailed description of the INMS instrument response under varying spacecraft conditions and laid the groundwork for follow-up studies focused solely on compositional analyses. However, early studies of the Enceladus plume---though foundational---may not have been well-suited to resolve minor species ambiguities for various reasons. In order to facilitate comparison with our methodology, we briefly describe the spectral deconvolution procedure developed by the authors of ref.~\cite{magee_inms-derived_2009} that has been implemented in various other works (e.g., refs.~\cite{waite_ion_2005, waite_process_2007, waite_chemical_2018}).

In their procedure, the authors first determine the contributions from major species through a visual analysis of prominent spectral features. Mixing ratios for the major species are estimated from the base peaks of each species, assuming they contribute 100\% of the measured counts at these mass channels. Minor species are then identified sequentially by subtracting their contributions from the total spectrum to produce a residual spectrum. For small portions of the spectrum where a few candidate species exhibit overlaying signatures, species are fit based on a custom-defined fit statistic using a grid search algorithm. Iterative minimization of the fit statistic is achieved numerically by sweeping through various mixing ratios at increasingly finer resolution. For species that share the same base peak (e.g., N$_2$ and C$_2$H$_4$), the fit statistic is manipulated to exclude this mass channel. The high computational intensity of the grid search algorithm prohibits fitting more than four species at a time. 

We believe that the methodology of ref.~\cite{magee_inms-derived_2009} described above, though useful and effective, may not be optimal for identifying minor species in the INMS data. The order in which species are subtracted from the initial spectrum will generally influence the outcome of the analysis (see the discrepancy between the exhaustive and forward modeling approaches presented in the main text). Although it is true that a bias-variance tradeoff can be useful for combatting high dimensionality, the procedure of ref.~\cite{magee_inms-derived_2009} is not amenable to quantitative assessments of inter-model uncertainty. Indeed, the authors note that subjectivity of their analysis is a valid concern. Furthermore, the practice of fitting individual species to small portions of the spectrum neglects potential contributions from complex compounds with cracking patterns that span a large mass range. Moreover, grid searches that consider only a few species at a time may not be able to reliably identify minor species when the spectral library is highly collinear. In this regime, covariances between mixing ratios become strongly model dependent (see Fig. 2 of the main text), and multi-model inference based on model-averaged parameters is warranted~\cite{richards_model_2011}. Additionally, the fit statistic described in ref.~\cite{magee_inms-derived_2009} is not a standard metric, and we suspect that the use of different fit statistics for different model parameters could lead to difficulties in interpreting the results. Lastly, the authors’ procedure does not employ dimensionality reduction or account for the possibility of over-fitting to noise. 

By contrast, our approach quantitatively addresses both inter- and intra-model uncertainty in the spectral decomposition of INMS data. While previous studies, such as those presented in refs.~\cite{postberg_plume_2018, mckinnon_mysterious_2018}, have concluded that minor species identification requires a higher resolution mass spectrometer, we have presented a mathematical framework capable of discriminating between previously ambiguous species. The heuristics used in this analysis are based on maximum likelihood estimation and relative entropy minimization—foundational principles of statistical inference and information theory. 

Nevertheless, this study is not without limitations. A major challenge for any compositional analysis of INMS data stems from the large number of candidate plume species. The chemistry of the ocean and ice shell could include hundreds to thousands of unique compounds that contribute to the observed INMS spectrum. Our approach using the AICc is based on the principle of parsimony in that the least-complex, best-fitting model is favored over similarly performing models of higher complexity. Although this is a fruitful approach to developing conservative models of the plume's composition, nature does not necessarily reflect this ideal. Future investigations using higher resolution mass spectrometers with larger mass ranges will shed light on the full extent of chemical diversity within the plume and the ocean beneath.

A number of interesting follow-up studies could be conducted to validate the results presented in this work. A strong approach would be to treat each of the slow flybys (E14, E17, and E18) as individual data sets, as opposed to averaging them together. Machine learning models could then be trained on one data set and evaluated on another. A sort of round-robin procedure could be used to estimate the uncertainty associated with training on a particular data set. Such a methodology would eliminate the need for heuristic statistics such as the AICc in favor of actual independent test set performance. This implementation would, however, require correcting for instrument artifacts in each of the individual Enceladus flybys.